\documentclass[11pt]{article}

\usepackage{hyperref}  
\usepackage{graphicx}  
\usepackage{amssymb}   
\usepackage{amsmath}   

\title{\bf Critical coupling and \ensuremath{\Delta\langle\phi^2\rangle} in three-dimensional \ensuremath{\phi^4} theory}

\author{Stephan D\"urr${}^{a,b}$ and Tolga S.\ H.\ Kiel${}^a$}

\date{${}^a${\sl Department of Physics, University of Wuppertal, 42119 Wuppertal, Germany}\\
${}^b${\sl J\"ulich Supercomputing Centre, Forschungszentrum J\"ulich, 52425 J\"ulich, Germany}}

\begin{document}

\maketitle

\begin{abstract}
We consider the $\phi^4$ theory with $\phi(x)\in\mathbb{R}$ in three Euclidean dimensions. For a variety of self-couplings $\hat{\lambda}$, the critical (bare) parameter $\hat{\mu}_{0\mathrm{c}}^2(\hat{\lambda})$ is determined where the lattice-regulated system changes from the symmetric phase to the broken phase. Next, we send the box volume to infinity and switch to a universal scheme; this yields $\hat{\mu}_\mathrm{c}^2(\hat{\lambda})$ for each simulated $\hat{\lambda}$. Finally, a continuum extrapolation gives $f_\mathrm{c}=\lim_{\hat{\lambda}\to0} \hat{\lambda}^2/\hat{\mu}_\mathrm{c}^2(\hat{\lambda})$, and we find $f_\mathrm{c} = 18.808(64)_{\mathrm{stat}}(45)_{\mathrm{sys}} = 18.808(78)_{\mathrm{tot}}$. In addition, $\Delta\langle\phi^2\rangle_\mathrm{c}/\lambda = -0.00303(12)_{\mathrm{stat}}(12)_{\mathrm{sys}} = -0.00303(17)_{\mathrm{tot}}$ is determined in the combined continuum and infinite-volume limit.
\end{abstract}


\section{Introduction\label{sec:intro}}

The $\phi^4$ theory in Euclidean space is defined by considering all field configurations $\phi(x)$ with $x\in\mathbb{R}^d$, and weighting them by $\exp(-S)$. We adopt the particle physics convention of using ``natural units'', i.e., all formulas are written modulo factors of $\hbar$, $k_\mathrm{B}$ and $c$. With these conventions, the Euclidean action takes the form
\begin{equation}
    S[\phi]=\int\mathrm{d}^dx\left\{\frac{1}{2}\partial_\rho\phi(x)\partial_\rho\phi(x)+\frac{\mu_0^2}{2}\phi(x)^2+\frac{\lambda_0}{4}\phi(x)^4\right\}
    \label{def_orig}
\end{equation}
where $d$ denotes the number of dimensions of spacetime. Requiring $S$ to be dimensionless gives $[\phi]=(d-2)/2$, where $[.]$ denotes the mass dimension of the argument. Specifically in $d=3$ dimensions (below abbreviated as ``3D''), this gives $[\phi]=1/2$, $[\mu_0^2]=2$, and $[\lambda_0]=1$. The coupling $\lambda$ thus has the dimension of a mass, and this means that the theory in 3D is super-renormalizable (that is, only a finite number of Feynman diagrams is primitively divergent); see any textbook on QFT.

It pays to have a look at the ``sister theories'' defined by the action (\ref{def_orig}) in 2D and 4D.
In 2D the coupling $\lambda$ has the dimension of a squared mass, i.e., the theory is super-renormalizable, too.
But here, things are even simpler than in 3D, since only one (non-vacuum) Feynman diagram is primitively divergent (the ``tadpole diagram''). As a result, in 2D only the mass parameter $\mu_0^2$ is renormalized into $\mu^2$, while setting $\lambda\equiv\lambda_0$ is a legitimate choice.
In 3D, once again, a finite number of (non-vacuum) Feynman diagrams (with up to two loops) is primitively divergent.
Upon matching to a standard continuum scheme (e.g.\ $\overline{\mathrm{MS}}$ at a given scale), only the continuum $\mu^2$ undergoes UV renormalization, while the continuum $\lambda$ receives a finite shift (see below).
In 4D, finally, the theory is critically renormalizable, since the coupling constant $\lambda$ is a pure number.
New divergences occur at any loop-order, and upon matching to the $\overline{\mathrm{MS}}$ scheme both $\mu^2$ and $\lambda$ undergo UV-renormalization (and field-renormalization $Z_\phi$ becomes an issue, too).
This theory has been found to be ``trivial'' in the sense that sending the regulator scale $\Lambda\to\infty$ drives the renormalized coupling $\lambda\to0$, so one ends off with a non-interacting theory \cite{Frohlich:1982tw,Aizenman:2019yuo}.
Despite this property the $\phi^4$ theory in 4D serves a nucleus of the Higgs sector in the Standard Model, with the proviso that it can only be an effective theory that is valid up to some (so-far inaccessible) energy scale, perhaps in the peta ($10^{15}$) or exa ($10^{18}$) electron-volt range.

We discretize the theory (in 3D) on a cubic lattice $\Lambda$ with spacing $a$ and periodic boundary conditions. The integral in (\ref{def_orig}) is replaced by $a^3\sum_x$, and the nearest-neighbor Laplacian (applied to an arbitrary function $f$ that lives on the lattice sites $x$) is
\begin{equation}
    (\triangle f)(x)=\frac{1}{a^2}\sum_{\rho=1:3} \left[f(x+ae_\rho)-2f(x)+f(x-ae_\rho)\right]
\end{equation}
where $e_\rho$ denotes the unit vector in the $\rho$-direction. The lattice action is therefore
\begin{equation}
    S=a^3\sum_{x\in\Lambda}\phi(x)\left(-\frac{1}{2}\triangle+\frac{\mu_0^2}{2}\right)\phi(x)+a^3\sum_{x\in\Lambda}\frac{\lambda_0}{4}\phi(x)^4
    \label{def_cont}
\end{equation}
where the field $\phi$ still has mass dimension $1/2$.
Since this is not practical on a computer, one tends to ``absorb'' a factor $a^{1/2}$ into $\phi$
(what used to be $a^{1/2}\phi$ is again denoted by $\phi$), and the action takes the form
\begin{equation}
    S=\sum_{x\in\Lambda}\left\{-\sum_{\rho=1:3}\phi(x)\phi(x+\hat{\rho})+\left(3+\frac{\hat{\mu}_0^2}{2}\right)\phi(x)^2+\frac{\hat{\lambda}_0}{4}\phi(x)^4\right\} \;.
    \label{def_final}
\end{equation}
Here $x+\hat{\rho}\equiv x+ae_\rho$ is the nearest-neighbor site in the $\rho$-direction, and the dimensionless parameters are
\begin{equation}
    \hat{\mu}_0^2=a^2\mu_0^2 \;,\qquad \hat{\lambda}_0=a\lambda_0.
\end{equation}

As mentioned previously, only the mass parameter requires ultraviolet (UV) renormalization in 3D. We define the continuum mass in the $\overline{\mathrm{MS}}$ scheme at the scale $\bar{\mu}=2\lambda$. The matched continuum coupling $\lambda$ does not require UV renormalization. But at any nonzero lattice spacing, it differs from the bare lattice coupling $\lambda_0$ by a finite amount. For the nearest-neighbor action in Eq.~\eqref{def_final}, the matching relations take the form~\cite{Arnold:2001ir,Sun:2002cc}
\begin{align}
    \hat{\mu}^2 = Z_{\hat{\mu}^2}^{-1}\hat{\mu}_0^2 - \delta_{\hat{\mu}^2} \quad \text{and}\quad{} \hat{\lambda}=\hat{\lambda}_0 - \delta_{\hat{\lambda}}\;,\label{eq:renormalization_relations}
\end{align}
with the multiplicative and additive matching/renormalization terms given by
\begin{equation}
    Z_{\hat{\mu}^2} = 1+3\frac{\xi}{4\pi}\hat{\lambda} \;,\quad 
    \delta_{\hat{\mu}^2} =- 3\frac{\Sigma}{4\pi}\hat{\lambda} + \frac{6}{(4\pi)^2} \left[C_3 + \ln\left(\frac{3}{\hat{\lambda}}\right) - 3\Sigma\xi\right]\hat{\lambda}^2 \;,\quad 
    \delta_{\hat{\lambda}} = 9\frac{\xi}{4\pi}\hat{\lambda}^2 \;.
\end{equation}
Specifically for our nearest-neighbor Laplacian, the finite coefficients read~\cite{Arnold:2001ir}
\begin{equation}
    \xi \simeq 0.152\;859\;324\;966\;101 \;,\quad 
    \Sigma \simeq 3.175\;911\;535\;625\;22 \;,\quad 
    C_3 \simeq 0.088\;480\;10 \;.
\end{equation}

Like in our previous study in 2D \cite{Durr:2025hfu}, we determine, for a given $\hat\lambda_0$ or $\hat\lambda$, the bare $\hat\mu_0^2$ where the system changes from the symmetric to the broken phase.
Based on the renormalized quantities $\hat{\lambda}$ and $\hat{\mu}_\mathrm{c}^2(\hat{\lambda})$ at the transition line, we extract the dimensionless critical coupling that is defined in 3D through the relation
\begin{equation}
    f_\mathrm{c} \equiv \lim_{\hat{\lambda}\to0} f_\mathrm{c}(\hat{\lambda}) \qquad\text{with}\qquad f_\mathrm{c}(\hat{\lambda})\equiv\frac{\hat{\lambda}^2}{\hat{\mu}_\mathrm{c}^2(\hat{\lambda})}
    \label{def_fc}
\end{equation}
while in 2D the last term was $f_\mathrm{c}(\hat\lambda_0)=\hat{\lambda}_0/\hat{\mu}_\mathrm{c}^2(\hat{\lambda}_0)$.
Note that $f_\mathrm{c}$ is independent of the lattice regularization to which the input parameter $\hat{\lambda}_0$ and the extracted value $\hat{\mu}_{0\mathrm{c}}^2$ would refer (provided that the continuum renormalization scheme and scale are held fixed).
In addition, the quantity $\Delta\langle\phi^2\rangle$ is defined as the difference between the expectation values of $\phi^2$ at the critical point in the interacting and free theories~\cite{Arnold:2001ir,Sun:2002cc}
\begin{equation}
    \Delta\langle\phi^2\rangle_\mathrm{c}(\lambda) \equiv [\langle\phi^2\rangle_\mathrm{c}]_\lambda-[\langle\phi^2\rangle_\mathrm{c}]_0 \;.
\end{equation}
For the nearest-neighbor lattice action (\ref{def_final}) the free critical value is $\Sigma/(4\pi)$, and a dimensionless ratio is
\begin{equation}
    \frac{\Delta\langle\phi^2\rangle_\mathrm{c}}{\lambda}
    \equiv\lim_{\hat{\lambda}\to0}
    \frac{\langle\phi^2\rangle_\mathrm{c}(\hat{\lambda})-\Sigma/(4\pi)}{\hat{\lambda}} \;.
    \label{eq:dphisq_lattice_limit}
\end{equation}
Note that the numerator and denominator on the right-hand side vanish as $\hat{\lambda}\to0$, but the ratio stays finite.
A different view on this is that both in \eqref{def_fc} and \eqref{eq:dphisq_lattice_limit} the continuum limit $a\to0$ is taken, while $\lambda$ is finite.
In this article, our objective is to obtain accurate numerical determinations of (\ref{def_fc}) and (\ref{eq:dphisq_lattice_limit}) based on precise Monte Carlo simulations.

The remainder of this paper is organized as follows.
In Sec.~\ref{sec:finite_vol} we specify the algorithms used and discuss how we scan, for a given set of parameters $(\hat{\lambda}_0,N_x)$, the bare $\hat{\mu}_0^2$ to locate the transition point. 
In Sec.~\ref{sec:inf_vol} we extrapolate our $\hat{\mu}_{0\mathrm{c}}^2$ and $\langle\phi^2\rangle_\mathrm{c}$ to infinite volume, match $\hat{\lambda}_0$ to $\hat{\lambda}$, convert $\hat{\mu}_{0\mathrm{c}}^2$ to $\hat{\mu}_\mathrm{c}^2$ to form $f_\mathrm{c}(\hat{\lambda})$ and $\Delta\langle\phi^2\rangle_\mathrm{c}(\hat{\lambda})/\hat{\lambda}$.
The most demanding part of the article is the analysis in Secs.~\ref{sec:cont_fc} and \ref{sec:cont_dphisq}, which yields $f_\mathrm{c}$ and $\Delta\langle\phi^2\rangle_\mathrm{c}/\lambda$ with good control over all statistical and systematic uncertainties involved.
Our final results are compared with previous determinations in Sec.~\ref{sec:conc}, while the finite-size scaling forms and the treatment of the uncertainties are reviewed in Apps.~\ref{sec:app_a} and \ref{sec:app_b}.
Finally, App.~\ref{sec:app_c} provides access to some fit details, and App.~\ref{sec:app_d} contains our primary data.


\section{Locating the phase transition point in finite volume\label{sec:finite_vol}}

We simulate the lattice action \eqref{def_final} on $N_x \times N_x \times N_x$ lattices with periodic boundary conditions in all directions. Generically in lattice field theory, there is an algorithmic issue in the vicinity of a phase transition. For the $\phi^4$ theory, an embedded cluster algorithm is useful to avoid critical slowing down~\cite{Brower:1989mt}. We combine several update algorithms and define an ``update package'' that consists of a multi-hit Metropolis sweep, six overrelaxation sweeps, one Hybrid Monte Carlo sweep, a global sign change of $\phi(x)$ across the lattice (performed with 50\% probability), a parity operation on $\phi(x)$ (again with 50\% probability), as well as two embedded single-cluster Wolff updates \cite{Wolff:1988uh}. We choose to separate two consecutive measurements by $n_\mathrm{sepa}=10$ such update packages. Consequently, our measurements are reasonably well decorrelated, but we process the data by means of a jackknife analysis with sufficiently large block size and determine, as a check, the normalized autocorrelation functions of our observables.

\begin{figure}[!tb]
    \includegraphics[width=0.5\linewidth]{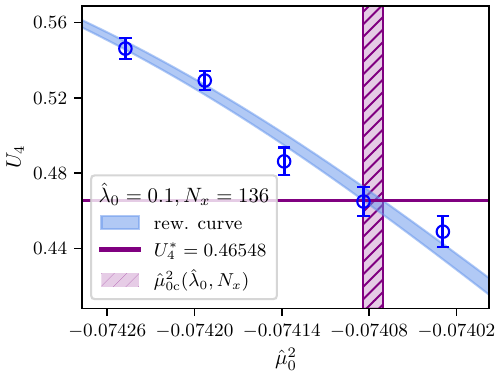}%
    \includegraphics[width=0.5\linewidth]{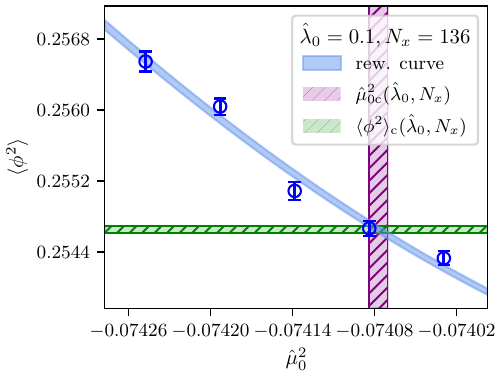}
    \caption{Left: The Binder cumulant $U_4$ versus $\hat{\mu}_0^2$ at $\hat{\lambda}_0=0.1$ for five independent simulations on $N_x=136$ lattices. The light blue band shows the combined (reweighted) curve with its $1\sigma$-uncertainty. The purple horizontal line reflects the universal value $U_4^*=0.46548$ from Ref.~\cite{Ferrenberg:2018zst}. The purple vertical band indicates $\hat{\mu}_{0\mathrm{c}}^2(\hat{\lambda}_0,N_x)$, i.e., the value of $\hat{\mu}_0^2$ at which the reweighted curve equals $U_4^*$; this defines the finite-volume transition. Right: The observable $\langle\phi^2\rangle$ versus $\hat{\mu}_0^2$, based on the same runs and reweighting procedure, with the same purple vertical band to indicate $\hat{\mu}_{0\mathrm{c}}^2(\hat{\lambda}_0,N_x)$. The green horizontal band shows the resulting $\langle\phi^2\rangle_\mathrm{c}(\hat{\lambda}_0,N_x)$, i.e., the reweighted curve evaluated at $\hat{\mu}_{0\mathrm{c}}^2(\hat{\lambda}_0,N_x)$.\label{fig:binder_phisq_reweighting}}
\end{figure}

For each pair $(\hat{\lambda}_0,\hat{\mu}_0^2)$ of bare parameters and a given box size $N_x$, we denote the volume-averaged field by $\bar{\phi}$ and define the Binder cumulant $U_4$ of this observable by~\cite{Binder:1981sa}
\begin{equation}
    U_4 = 1-\frac{\langle\bar{\phi}^{\,4}\rangle}{3\langle\bar{\phi}^{\,2}\rangle^2} \qquad \mbox{with} \qquad \bar{\phi}\equiv\frac{1}{N_x^3}\sum_{x\in\Lambda}\phi(x) \;.
    \label{eq:binder_cumulant_def}
\end{equation}
The second observable of interest is the expectation value of the volume average of the squared field
\begin{equation}
    \langle\phi^2\rangle\equiv\langle\overline{\phi^2}\rangle
    \qquad \mbox{with} \qquad
    \overline{\phi^2}\equiv\frac{1}{N_x^3}\sum_{x\in\Lambda}\phi(x)^2 \;.
    \label{eq:phisq_observable_def}
\end{equation}

\begin{figure}[!tb]
    \includegraphics[width=0.5\linewidth]{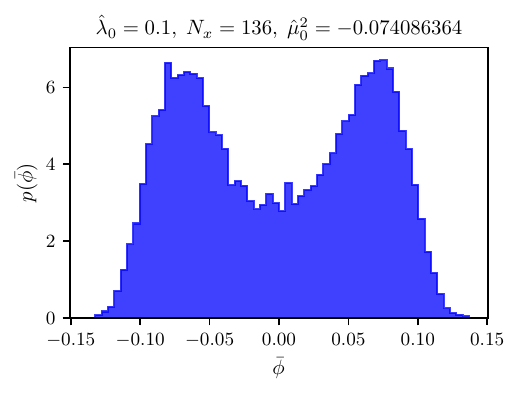}%
    \includegraphics[width=0.5\linewidth]{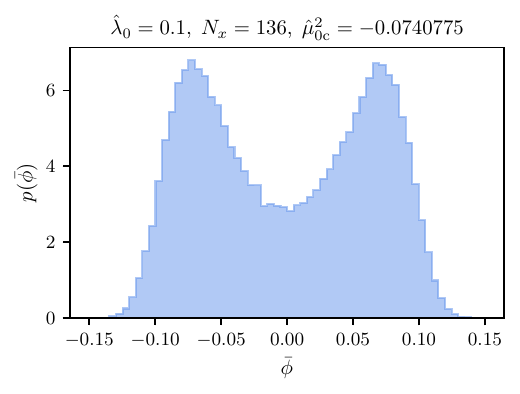}
    \caption{\label{fig:phi_histogram_reweighting_check}Probability distribution of the volume-averaged field $\bar{\phi}$ ``at the transition point'' for $\hat{\lambda}_0=0.1$ and $N_x=136$. Left: distribution for the simulated ensemble at $\hat{\mu}_0^2=-0.074086364$, which is the one closest to $\hat{\mu}_{0\mathrm{c}}^2(0.1,136)$. Right: combined distribution at the critical value $\hat{\mu}_{0\mathrm{c}}^2(0.1,136)=-0.0740775$, i.e., each ensemble shown in Fig.~\ref{fig:binder_phisq_reweighting} is reweighted to $\hat{\mu}_{0\mathrm{c}}^2$, every histogram bin is jackknifed separately, and the ensemble estimates in each bin are combined using inverse-variance weights.}
\end{figure}

There are two options for using the Binder cumulant to locate the phase transition point $\hat{\mu}_{0\mathrm{c}}^2$ in a finite volume. In the conventional crossing method, one determines $U_4$ as a function of $\hat{\mu}_0^2$ for one $\hat{\lambda}_0$ and a pair of box sizes $N_x$. The intersection of the two curves defines $\hat{\mu}_{0\mathrm{c}}^2(\hat{\lambda}_0,N_x^\mathrm{min},N_x^\mathrm{maj})$ for the given $\hat{\lambda}_0$ and the given $N_x$-pair. The result is attributed to an $N_x$ anywhere in the range $N_x^\mathrm{min}<N_x<N_x^\mathrm{maj}$, and finally the infinite-volume limit is taken to find $\hat{\mu}_{0\mathrm{c}}^2$ for the given $\hat{\lambda}_0$~\cite{Binder:1981sa}. In this limit the value of $U_4$ at the intersection approaches the universal value $U_4^*$. With prior knowledge of $U_4^*$ in hand, one can instead use the fixed-point method in which, for given $\hat{\lambda}_0$ and $N_x$, the point $\hat{\mu}_0^2$ where $U_4$ reaches $U_4^*$ is determined~\cite{Sun:2002cc,Campostrini:2006ms}. This yields $\hat{\mu}_{0\mathrm{c}}^2(\hat{\lambda}_0,N_x)$, and again an infinite-volume extrapolation is needed to find $\hat{\mu}_{0\mathrm{c}}^2(\hat{\lambda}_0)$ for the given $\hat{\lambda}_0$. We use the second method, because it yields smaller statistical uncertainties in our analysis.

The value of $U_4^*$ to be used depends on the universality class of the model. The scalar $\phi^4$ theory in 3D is generally accepted to be in the 3D Ising universality class (see the tables in Sec.\,3.2.1 of Ref.~\cite{Pelissetto:2000ek}), and we use the 3D Ising value $U_4^*=0.46548(5)$ from Ref.~\cite{Ferrenberg:2018zst}. Furthermore, the pseudo-critical mass parameter $\hat{\mu}_{0\mathrm{c}}^2(\hat{\lambda}_0,N_x)$ is known to approach its infinite-volume value with asymptotic corrections of the form $N_x^{-1/\nu-\omega}$; the finite-size scaling argument is reviewed in App.~\ref{sec:app_a}.

For a given $(\hat{\lambda}_0,N_x)$ combination, we simulate several values of $\hat{\mu}_0^2$ in the vicinity of the transition point. Each ensemble is reweighted with the Ferrenberg-Swendsen method~\cite{Ferrenberg:1988yz,Ferrenberg:1989ui}, and the resulting curves are combined with the variance-weighted prescription of Holm and Janke~\cite{Holm:1993zz}. As mentioned before, we use the central value $U_4^*=0.46548$ from Ref.~\cite{Ferrenberg:2018zst}. Repeating our analysis with $U_4^*$ shifted by up to $\pm3\sigma$ changes $\hat{\mu}_{0\mathrm{c}}^2$ by about one tenth of its statistical uncertainty. Hence, the uncertainty due to the error-bar of the input quantity $U_4^*$ is negligible.

In the left panel of Fig.~\ref{fig:binder_phisq_reweighting} an example of the Binder cumulant analysis for $\bar{\phi}$ is shown (here at $\hat{\lambda}_0=0.1$ and $N_x=136$). The five independent simulations (blue circles) are separately reweighted and then combined (light blue band). The point where the combined curve reaches $U_4^*$ (purple horizontal line) determines $\hat{\mu}_{0\mathrm{c}}^2(\hat{\lambda}_0,N_x)$ (purple vertical band). The expectation value of the second observable, $\langle\phi^2\rangle$, is reweighted with the same Monte Carlo histories. Its value at the pseudo-critical coupling $\hat{\mu}_{0\mathrm{c}}^2(\hat{\lambda}_0,N_x)$ defines $\langle\phi^2\rangle_\mathrm{c}(\hat{\lambda}_0,N_x)$ (green band), and this procedure is shown in the right panel.

As a consistency check, we illustrate in Fig.~\ref{fig:phi_histogram_reweighting_check} the effect that the reweighting procedure has on the histogram of $\bar{\phi}$ as defined in (\ref{eq:binder_cumulant_def}). The left panel shows the histogram in the simulation that is closest (in $\hat{\mu}_0^2$) to the transition point $\hat{\mu}_{0\mathrm{c}}^2(0.1,136)$. One observes a clear two peak structure, but the weight of the two peaks is not exactly identical. In the right panel, the situation is shown after reweighting. This time, the two peaks have the same height (within errors), so the reweighting procedure seems to work fine.

We repeat this analysis for a set of $16$ values of $\hat{\lambda}_0$ between $0.07$ and $1.8$. For each of them, a range of box sizes $N_x$ is chosen that is large enough to allow for an extrapolation $N_x\to\infty$ as described in the next section. The finite-volume values of $\hat{\mu}_{0\mathrm{c}}^2$ and $\langle\phi^2\rangle_\mathrm{c}$ for every simulated combination $(\hat{\lambda}_0,N_x)$ are collected in App.~\ref{sec:app_d}.


\section{Extrapolation to infinite volume\label{sec:inf_vol}}

With the pseudocritical values $\hat{\mu}_{0\mathrm{c}}^2(\hat{\lambda}_0,N_x)$ and $\langle\phi^2\rangle_\mathrm{c}(\hat{\lambda}_0,N_x)$ in hand, we remove finite-volume effects by taking the limit $N_x\to\infty$. This step requires knowledge of the functional form of the finite-size corrections. In Ref.~\cite{Sun:2002cc}, the leading finite-size dependence was first determined and then used to correct the result of the continuum extrapolation, which was taken in constant physical volume. Here we take the infinite-volume limit separately for each $\hat{\lambda}_0$, and then take the continuum limit in a second step. To judge the quality of either extrapolation, we vary the number of correction terms in the fit ansatz, as well as the range to be included in the fit. Finally, the acceptable fits are combined according to their AIC weights (see below).

\begin{figure}[!tb]
    \includegraphics[width=0.5\linewidth]{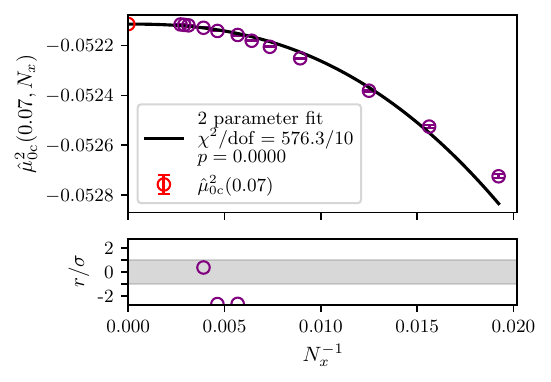}%
    \includegraphics[width=0.5\linewidth]{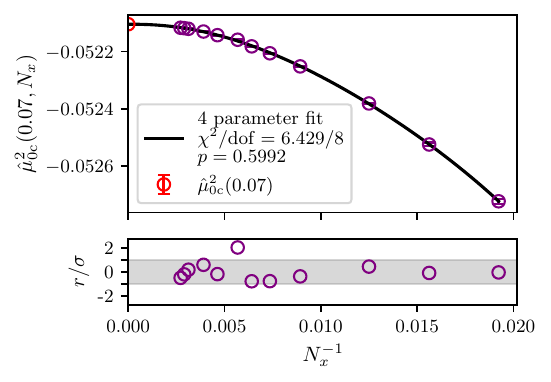}
    \caption{\label{fig:fss_musq}Two examples of the infinite-volume extrapolation of $\hat{\mu}_{0\mathrm{c}}^2(0.07,N_x)$, i.e., at our smallest bare coupling $\hat\lambda_0$. In both cases all available $N_x$ are included. Left: leading-order ansatz ($n_\mathrm{par}=2$ in Eq.~\ref{eq:mu0sq_fit_ansatz}), resulting in $p=0.0000$ at the displayed precision. Right: ansatz with two subleading terms ($n_\mathrm{par}=4$ in Eq.~\ref{eq:mu0sq_fit_ansatz_4_param}), giving $p=0.5992$. The sub-panels show the fit residuals (data-fit) in units of the statistical uncertainties of the data.}
\end{figure}

Let us first consider the leading-order finite-size scaling ansatz~\cite{Sun:2002cc}
\begin{align}
    \hat{\mu}_{0\mathrm{c}}^2(\hat{\lambda}_0,N_x)=\hat{\mu}_{0\mathrm{c}}^2(\hat{\lambda}_0)+c_1N_x^{-1/\nu-\omega}
    \label{eq:mu0sq_fit_ansatz}
\end{align}
with the exponents fixed%
\footnote{Varying $\omega$ and $\nu$ independently by up to $\pm3\sigma$ changes $\hat{\mu}_{0\mathrm{c}}^2(\hat{\lambda}_0)$ by less than one-tenth of the statistical uncertainty.}
to the central values of $\nu=0.629\,912(86)$~\cite{Ferrenberg:2018zst} and $\omega=0.8303(18)$~\cite{Lundow:2017xpf}.
A typical extrapolation of $\hat{\mu}_{0\mathrm{c}}^2(\hat{\lambda}_0=0.07,N_x)$ in $1/N_x$ is shown in the left panel of Fig.~\ref{fig:fss_musq}.
Including all lattice sizes gives $\hat{\mu}_{0\mathrm{c}}^2(0.07)=-0.05211479(43)$ with $\chi^2/\mathrm{dof}=576.3/10$; so the leading-order ansatz does not adequately describe the full range of box volumes. 

\begin{figure}[!tb]
    \includegraphics[width=\linewidth]{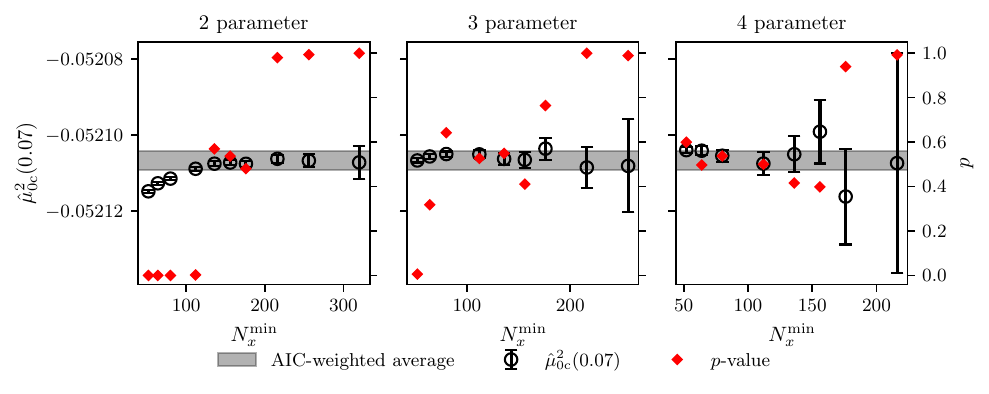}
    \caption{\label{fig:fss_musq_scan_combined}Extracted $\hat{\mu}_{0\mathrm{c}}^2(0.07)$ values in infinite volume (black circles, left axis) and the $p$-value of the respective fit (red diamonds, right axis) as a function of $N_x^{\min}$. The three panels refer to $n_\mathrm{par}=2$ (left), 3 (center), and 4 (right) in Eq.~\ref{eq:mu0sq_fit_ansatz_4_param}. The horizontal band indicates the final AIC-weighted combination over all fits with $p>0.05$.} 
\end{figure}

The extrapolated $\hat{\mu}_{0\mathrm{c}}^2(\hat{\lambda}_0)$ and the fit quality $p$ change if some of the smaller $N_x$ are excluded from the fit. The left panel of Fig.~\ref{fig:fss_musq_scan_combined} shows the extrapolated result and the $p$-value as a function of $N_x^\mathrm{min}$. The leading-order ansatz ($n_\mathrm{par}=2$) gives an acceptable fit only for $N_x^\mathrm{min}\geq136$, where the number of degrees of freedom that remain is quite limited. The result is stable in this region, but there is no unique choice of $N_x^\mathrm{min}$.

We choose to extend the ansatz of Eq.~\eqref{eq:mu0sq_fit_ansatz} by including one or two subleading corrections to the leading-order scaling behavior, that is, we also consider the functional forms
\begin{align}
    \hat{\mu}_{0\mathrm{c}}^2(\hat{\lambda}_0,N_x)&=\hat{\mu}_{0\mathrm c}^2(\hat{\lambda}_0)+c_1N_x^{-1/\nu-\omega}+c_2N_x^{-1/\nu-2\omega} \;,
    \label{eq:mu0sq_fit_ansatz_3_param}\\
    \hat{\mu}_{0\mathrm{c}}^2(\hat{\lambda}_0,N_x)&=\hat{\mu}_{0\mathrm c}^2(\hat{\lambda}_0)+c_1N_x^{-1/\nu-\omega}+c_2N_x^{-1/\nu-2\omega}+c_3N_x^{-1/\nu-3\omega}
    \label{eq:mu0sq_fit_ansatz_4_param}
\end{align}
with $n_\mathrm{par}=3$ or $n_\mathrm{par}=4$ free parameters, respectively (see App.~\ref{sec:app_a} for details).

The right panel of Fig.~\ref{fig:fss_musq} displays the four-parameter fit with all box sizes included; it gives $p=0.5992$, and the pulls show no obvious trend.
In Fig.~\ref{fig:fss_musq_scan_combined} the extrapolated $\hat{\mu}_{0\mathrm{c}}^2$ and $p$-values are shown for $n_\mathrm{par}=2,3,4$ and various $N_x^\mathrm{min}$. For $n_\mathrm{par}=3$ the result is stable as soon as the three smallest $N_x$ are excluded. The four-parameter ansatz works for the full range of box sizes, with essentially no residual dependence on $N_x^\mathrm{min}$. To aggregate all viable fits into a final result, we keep all fits with $p>0.05$. For each fit, the AIC-based weight~\cite{Akaike:1974vps,BMW:2014pzb,Borsanyi:2020mff}
\begin{equation}
    w_\mathrm{AIC}=\exp\left(-\frac{1}{2}\left[\chi^2+2n_\mathrm{par}-n_\mathrm{data}\right]\right) \label{eq:w_aic}
\end{equation}
is computed, where $n_\mathrm{data}$ is the number of data points and $n_\mathrm{par}$ the number of fit parameters. The term $-n_\mathrm{data}$ follows the recommendation of Refs.~\cite{BMW:2014pzb,Borsanyi:2020mff} on how to combine several fits with different $n_\mathrm{dof}=n_\mathrm{data}-n_\mathrm{par}$. In the end, the weights are normalized so that $\sum_i w_{\mathrm{AIC},i}=1$. The central value of a given observable is its AIC-weighted average, and its statistical uncertainty is obtained by repeating the analysis on every jackknife sample. The result for $\hat{\mu}_{0\mathrm{c}}^2(\hat{\lambda}_0=0.07)$ is shown as the gray horizontal band in Fig.~\ref{fig:fss_musq_scan_combined}.

\begin{figure}[!tb]
    \includegraphics[width=0.5\linewidth]{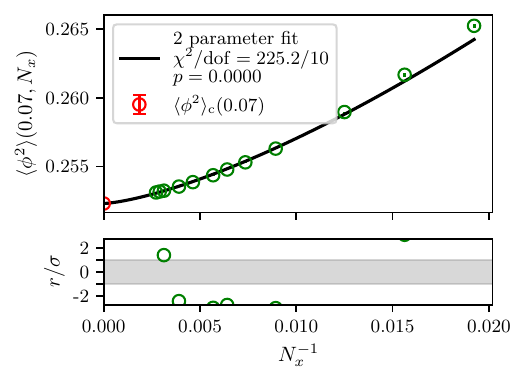}%
    \includegraphics[width=0.5\linewidth]{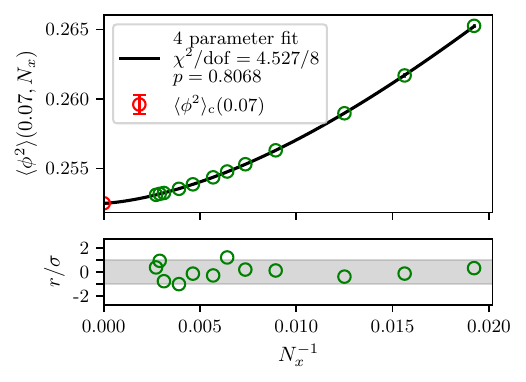}
    \caption{\label{fig:fss_phisq}Two examples of the infinite-volume extrapolation of $\langle\phi^2\rangle_\mathrm{c}(0.07,N_x)$ at our smallest bare coupling $\hat\lambda_0$. In both cases all available $N_x$ are included. Left: Leading-order ansatz ($n_\mathrm{par}=2$ in Eq.~\ref{eq:phisq_fit_ansatz}); the $p$-value is zero at the displayed precision. Right: Ansatz with two subleading correction terms ($n_\mathrm{par}=4$ in Eq.~\ref{eq:phisq_fit_ansatz}), giving $p=0.8068$. The lower panels show the pulls (data-fit) in units of the statistical uncertainties.}
\end{figure}

\begin{figure}[!tb]
    \includegraphics[width=\linewidth]{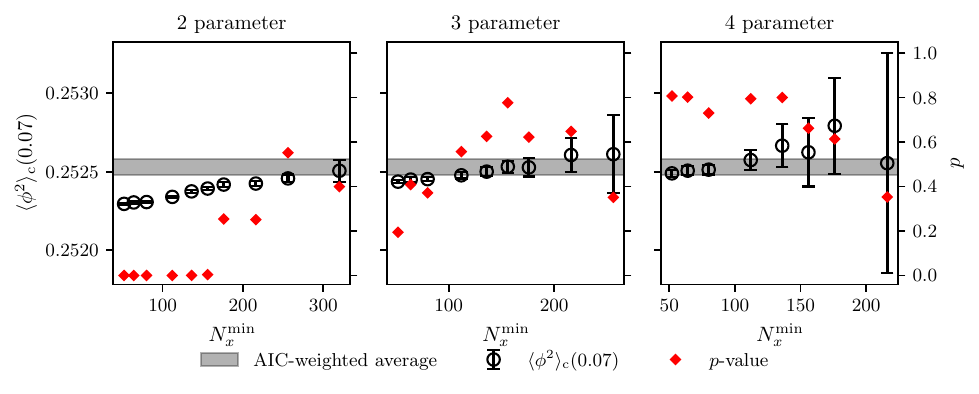}
    \caption{\label{fig:fss_phisq_scan_combined}Extracted $\langle\phi^2\rangle_\mathrm{c}(0.07)$ in infinite volume (black circles, left axis) and the $p$-value of the respective fit (red diamonds, right axis) as a function of $N_x^{\min}$. The three panels refer to $n_\mathrm{par}=2$ (left), 3 (center), and 4 (right) in Eq.~\ref{eq:phisq_fit_ansatz}. The horizontal band denotes the final AIC-weighted combination over all fits with $p>0.05$.} 
\end{figure}

To extrapolate the second observable, $\langle\phi^2\rangle_\mathrm{c}$, we use the general finite-size scaling ansatz
\begin{equation}
    \langle\phi^2\rangle_\mathrm{c}(\hat{\lambda}_0,N_x)=\langle\phi^2\rangle_\mathrm{c}(\hat{\lambda}_0)+c_1N_x^{-3+1/\nu}+c_2N_x^{-3/2-\omega}+c_3N_x^{-3/2-2\omega}
    \label{eq:phisq_fit_ansatz}
\end{equation}
that uses $n_{\mathrm{par}}=4$ (see App.~\ref{sec:app_a} for details). Setting $c_3=0$ gives $n_\mathrm{par}=3$, and $c_2=c_3=0$ means $n_\mathrm{par}=2$. Two fits that include all $N_x$ for $\hat{\lambda}_0=0.07$ are shown in Fig.~\ref{fig:fss_phisq}, and the scans versus $N_x^\mathrm{min}$ are assembled in Fig.~\ref{fig:fss_phisq_scan_combined}. In the latter case, the overall picture is similar to Fig.~\ref{fig:fss_musq_scan_combined}. The AIC-weighted average is found to be a robust and reliable estimate for $\langle\phi^2\rangle_\mathrm{c}(0.07)$; it is displayed by the gray horizontal band in Fig.~\ref{fig:fss_phisq_scan_combined}.

\begin{table}[!tb]
    \centering
    \begin{tabular}{@{\hskip 5pt}c@{\hskip 5pt}c@{\hskip 5pt}c@{\hskip 5pt}c@{\hskip 5pt}c@{\hskip 5pt}c@{\hskip 5pt}c@{\hskip 5pt}c@{\hskip 5pt}}
    \hline
    $\hat{\lambda}_0$ & $\hat{\lambda}$ & $\hat{\mu}_{0\mathrm{c}}^2$ & $\hat{\mu}_{\mathrm{c}}^2$ & $f_\mathrm{c}(\hat{\lambda})$ & $\langle\phi^2\rangle_\mathrm{c}$ & $\Delta\langle\phi^2\rangle_\mathrm{c}$ & $\Delta\langle\phi^2\rangle_\mathrm{c}/\hat{\lambda}$ \\
    \hline
    1.80 & $1.540272$ & $-1.135857(13)$ & $0.155627(12)$ & $15.2444(12)$ & $0.249019(34)$ & $-0.003712(34)$ & $-0.002410(22)$ \\
    1.50 & $1.311652$ & $-0.964239(22)$ & $0.109630(21)$ & $15.6931(30)$ & $0.249410(33)$ & $-0.003321(33)$ & $-0.002532(25)$ \\
    1.20 & $1.073774$ & $-0.7876106(86)$ & $0.0711319(83)$ & $16.2092(19)$ & $0.249958(31)$ & $-0.002773(31)$ & $-0.002583(29)$ \\
    1.00 & $0.909451$ & $-0.666571(13)$ & $0.049860(13)$ & $16.5884(42)$ & $0.250336(43)$ & $-0.002395(43)$ & $-0.002633(47)$ \\
    0.90 & $0.825412$ & $-0.604936(14)$ & $0.040579(14)$ & $16.7896(58)$ & $0.250553(48)$ & $-0.002178(48)$ & $-0.002639(58)$ \\
    0.80 & $0.740043$ & $-0.5424945(90)$ & $0.0322060(88)$ & $17.0050(46)$ & $0.250726(45)$ & $-0.002005(45)$ & $-0.002709(61)$ \\
    0.70 & $0.653278$ & $-0.479145(10)$ & $0.0247862(98)$ & $17.2181(68)$ & $0.250892(34)$ & $-0.001839(34)$ & $-0.002815(52)$ \\
    0.60 & $0.565046$ & $-0.414860(17)$ & $0.018278(17)$ & $17.468(16)$ & $0.25122(10)$ & $-0.00152(10)$ & $-0.00268(18)$ \\
    0.50 & $0.475271$ & $-0.3494641(62)$ & $0.0127678(61)$ & $17.6916(84)$ & $0.251334(30)$ & $-0.001397(30)$ & $-0.002940(63)$ \\
    0.40 & $0.383868$ & $-0.2828636(47)$ & $0.0082374(46)$ & $17.889(10)$ & $0.251599(66)$ & $-0.001132(66)$ & $-0.00295(17)$ \\
    0.30 & $0.290746$ & $-0.2149239(80)$ & $0.0046711(79)$ & $18.097(31)$ & $0.251964(77)$ & $-0.000767(77)$ & $-0.00264(27)$ \\
    0.25 & $0.243508$ & $-0.1803969(55)$ & $0.0032426(55)$ & $18.287(31)$ & $0.252074(56)$ & $-0.000657(56)$ & $-0.00270(23)$ \\
    0.20 & $0.195803$ & $-0.1454138(67)$ & $0.0020911(66)$ & $18.334(58)$ & $0.252156(50)$ & $-0.000575(50)$ & $-0.00293(25)$ \\
    0.15 & $0.147614$ & $-0.1099681(20)$ & $0.0011804(20)$ & $18.459(32)$ & $0.252375(63)$ & $-0.000356(63)$ & $-0.00241(43)$ \\
    0.10 & $0.098929$ & $-0.0739857(15)$ & $0.0005272(15)$ & $18.562(53)$ & $0.252412(30)$ & $-0.000319(30)$ & $-0.00323(31)$ \\
    0.07 & $0.069472$ & $-0.0521066(24)$ & $0.0002584(24)$ & $18.68(17)$ & $0.252530(51)$ & $-0.000201(51)$ & $-0.00289(73)$ \\
    \hline
    \end{tabular}
    \caption{\label{tab:infinite_volume_results}Infinite-volume results at each simulated value of the bare coupling $\hat{\lambda}_0$. The subsequent columns list the renormalized coupling $\hat{\lambda}$, the infinite volume bare and renormalized mass parameters $\hat{\mu}_{0\mathrm{c}}^2$ and $\hat{\mu}_\mathrm{c}^2$, the ratio $f_\mathrm{c}(\hat{\lambda})=\hat{\lambda}^2/\hat{\mu}_\mathrm{c}^2$, the infinite volume $\langle\phi^2\rangle_\mathrm{c}$ and its shifted counterpart $\Delta\langle\phi^2\rangle_\mathrm{c}=\langle\phi^2\rangle_\mathrm{c}-\Sigma/4\pi$, as well as $\Delta\langle\phi^2\rangle_\mathrm{c}/\hat{\lambda}$. These are the inputs to the continuum extrapolations described in Secs.~\ref{sec:cont_fc} and \ref{sec:cont_dphisq}, respectively.}
\end{table}

Finally, this procedure is repeated for all available $\hat{\lambda}_0$. The infinite-volume values of $\hat{\mu}_{0\mathrm{c}}^2$ and $\langle\phi^2\rangle_\mathrm{c}$ are listed in the third and sixth columns of Tab.~\ref{tab:infinite_volume_results}, respectively. Solving Eq.~\ref{eq:renormalization_relations} gives the renormalized $\hat{\lambda}$ and $\hat{\mu}_\mathrm{c}^2$ in the second and fourth columns. Finally, the fifth column follows from $f_\mathrm{c}(\hat{\lambda})=\hat{\lambda}^2/\hat{\mu}_\mathrm{c}^2(\hat{\lambda})$, while the last two columns contain $\Delta\langle\phi^2\rangle_\mathrm{c}$ and $\Delta\langle\phi^2\rangle_\mathrm{c}/\hat{\lambda}$ as used on the right-hand side of \eqref{eq:dphisq_lattice_limit}, i.e., before the continuum limit is taken.


\section{Continuum limit of \ensuremath{f_\mathrm{c}}\label{sec:cont_fc}}

From a look at Tab.~\ref{tab:infinite_volume_results} it is clear that extrapolating $f_\mathrm{c}(\hat{\lambda})$ to $\hat{\lambda}=0$ is non-trivial. The reason is the steady increase of the statistical uncertainty towards the continuum ($\hat{\lambda}=0$); it grows by two orders of magnitude from $0.0012$ at our largest coupling to $0.17$ at the smallest one. Thus, a fit over the full range will unavoidably be dominated by our precise data at large $\hat{\lambda}$, while the (noisy) data at small $\hat{\lambda}$ should be the most relevant ones in view of the desired limit.

In addition, there is no a-priori functional form for fitting the data. In Ref.~\cite{Arnold:2001ir} the data for the 3D model with $N=2$ (see below) were fitted with a linear ansatz, and the same choice was made for the 3D model with $N=1$ in Ref.~\cite{Sun:2002cc}. In two dimensions, people started fitting their data with linear ans\"atze and low-order polynomials~\cite{Loinaz:1997az}, but soon it was realized that more precise data contain logarithmic contributions~\cite{Schaich:2009jk}. Our 2D analysis showed that any unduly restriction on the functional basis would bias the result, even if some of the simpler fits look plausible~\cite{Durr:2025hfu}. This motivates us to include logarithmic terms in the present analysis, too. In the absence of further analytical guidance, we consider the same family of continuum extrapolations
\begin{equation}
f_\mathrm{c}(\hat{\lambda})=c_0+c_1\hat{\lambda}+c_2\hat{\lambda}\ln(\hat{\lambda})+c_3\hat{\lambda}\ln^2(\hat{\lambda})+c_4\hat{\lambda}^2\ln(\hat{\lambda})+c_5\hat{\lambda}^2\ln^2(\hat{\lambda})+c_6\hat{\lambda}^2
\label{eq:fc_cont_fit}
\end{equation}
that we used in~\cite{Durr:2025hfu}. This is a restricted version of $\sum_{m,n\geq0}c_{mn}\hat{\lambda}^m\ln^n(\hat{\lambda})$, where combinations with $(m=0,n>0)$ are prohibited, to ensure a finite limit $\hat{\lambda}\to0$. Beyond this, we are unaware of a physics based argument why one should exclude one of the terms in Eq.~\eqref{eq:fc_cont_fit}. The continuum value in each fit is $f_\mathrm{c}=c_0$.

\begin{figure}[!tb]
    \includegraphics[width=0.5\textwidth]{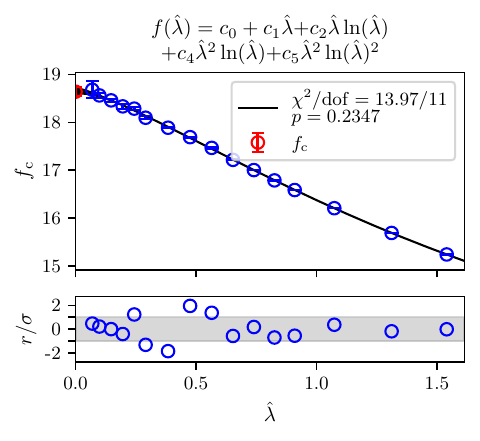}%
    \includegraphics[width=0.5\textwidth]{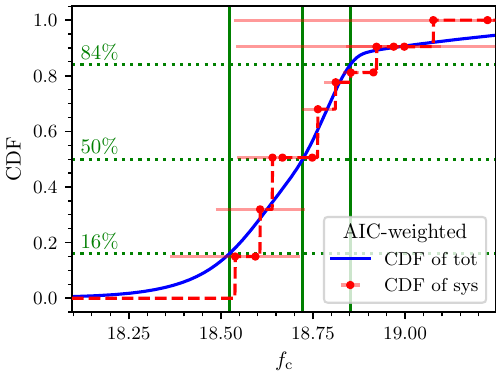}
    \caption{\label{fig:fc_lambdamax_1.8}
    Left: Example of a continuum extrapolation of $f_\mathrm{c}(\hat{\lambda})$ with a $\hat\lambda^\mathrm{max}$ that corresponds to $\hat{\lambda}_0^\mathrm{max}=1.8$, based on the ansatz with $\{c_2,c_4,c_5\}$. Right: Cumulative distribution function (CDF) of all statistically acceptable fits with $\hat{\lambda}_0^\mathrm{max}=1.8$, weighted by their AIC scores. At each step of the red dashed curve, the light-red horizontal band indicates the statistical uncertainty of that fit. The blue curve is the CDF obtained by combining the AIC weights with the Gaussian distributions of the statistical uncertainties. This yields $f_\mathrm{c}=18.72(11)_\mathrm{stat}(12)_\mathrm{sys}=18.72(16)_\mathrm{tot}$ for $\hat{\lambda}_0^\mathrm{max}=1.8$.}
\end{figure}

The goal is thus to determine which terms with $\{c_2,c_3,c_4,c_5,c_6\}$ are needed for a proper representation of the data (the linear part with $\{c_0,c_1\}$ is always needed). We begin our discussion by including all $c_i$ and the complete set of $\hat{\lambda}_0$, so $\hat{\lambda}_0^\mathrm{max}=1.8$. A representative fit over this range is shown in the left panel of Fig.~\ref{fig:fc_lambdamax_1.8}. Note that the fit itself uses the renormalized $\hat{\lambda}$, but it is easier to quote $\hat{\lambda}_0^\mathrm{max}$. The specific fit shown is the result of a pruning procedure. Starting from the linear ansatz, we add each of the five remaining terms separately and keep these fits. The fit with the highest $w_\mathrm{AIC}$ is selected for further refinement and is augmented by all possible terms that are not yet present. Again, all resulting fits are kept, but only the one with the largest $w_\mathrm{AIC}$ is used as a starting point for further refinement. The procedure stops when none of the new fits has a larger $w_\mathrm{AIC}$ than the parent fit.

\begin{table}[!tb]
\centering
\begin{tabular}{@{\hskip 5pt}l@{\hskip 1pt}c@{\hskip 5pt}c@{\hskip 5pt}c@{\hskip 5pt}c@{\hskip 5pt}}
\hline
fit ansatz  &  $f_\mathrm{c}$  &  $\chi^2/\mathrm{dof}$  &  $p$  &  $w_\mathrm{AIC}$  \\
\hline
$f(\hat{\lambda})=c_0+c_1\hat{\lambda}$ & $18.5939(41)$ & 1454/14 & 0.0000 & 0.0000 \\
\hline
$f(\hat{\lambda})=c_0+c_1\hat{\lambda}+c_2\hat{\lambda}\ln(\hat{\lambda})$ & $19.224(17)$ & 160.1/13 & 0.0000 & 0.0000 \\
$f(\hat{\lambda})=c_0+c_1\hat{\lambda}+c_3\hat{\lambda}\ln(\hat{\lambda})^2$ & $18.6673(43)$ & 34.05/13 & 0.0012 & 0.0000 \\
$f(\hat{\lambda})=c_0+c_1\hat{\lambda}+c_4\hat{\lambda}^2\ln(\hat{\lambda})$ & $18.9142(87)$ & 29.42/13 & 0.0057 & 0.0001 \\
$f(\hat{\lambda})=c_0+c_1\hat{\lambda}+c_5\hat{\lambda}^2\ln(\hat{\lambda})^2$ & $18.7481(54)$ & 44.98/13 & 0.0000 & 0.0000 \\
$f(\hat{\lambda})=c_0+c_1\hat{\lambda}+c_6\hat{\lambda}^2$ & $18.998(11)$ & 59.60/13 & 0.0000 & 0.0000 \\
\hline
$f(\hat{\lambda})=c_0+c_1\hat{\lambda}+c_2\hat{\lambda}\ln(\hat{\lambda})+c_3\hat{\lambda}\ln(\hat{\lambda})^2$ & $18.347(77)$ & 15.83/12 & 0.1992 & 0.0200 \\
$f(\hat{\lambda})=c_0+c_1\hat{\lambda}+c_2\hat{\lambda}\ln(\hat{\lambda})+c_4\hat{\lambda}^2\ln(\hat{\lambda})$ & $18.763(43)$ & 16.10/12 & 0.1867 & 0.0174 \\
$f(\hat{\lambda})=c_0+c_1\hat{\lambda}+c_2\hat{\lambda}\ln(\hat{\lambda})+c_5\hat{\lambda}^2\ln(\hat{\lambda})^2$ & $18.901(33)$ & 22.65/12 & 0.0308 & 0.0007 \\
$f(\hat{\lambda})=c_0+c_1\hat{\lambda}+c_2\hat{\lambda}\ln(\hat{\lambda})+c_6\hat{\lambda}^2$ & $18.684(49)$ & 14.65/12 & 0.2613 & 0.0360 \\
$f(\hat{\lambda})=c_0+c_1\hat{\lambda}+c_3\hat{\lambda}\ln(\hat{\lambda})^2+c_4\hat{\lambda}^2\ln(\hat{\lambda})$ & $18.97(13)$ & 29.26/12 & 0.0036 & 0.0000 \\
$f(\hat{\lambda})=c_0+c_1\hat{\lambda}+c_3\hat{\lambda}\ln(\hat{\lambda})^2+c_5\hat{\lambda}^2\ln(\hat{\lambda})^2$ & $18.7046(98)$ & 16.66/12 & 0.1626 & 0.0131 \\
$f(\hat{\lambda})=c_0+c_1\hat{\lambda}+c_3\hat{\lambda}\ln(\hat{\lambda})^2+c_6\hat{\lambda}^2$ & $18.19(13)$ & 20.65/12 & 0.0557 & 0.0018 \\
$f(\hat{\lambda})=c_0+c_1\hat{\lambda}+c_4\hat{\lambda}^2\ln(\hat{\lambda})+c_5\hat{\lambda}^2\ln(\hat{\lambda})^2$ & $18.853(21)$ & 19.32/12 & 0.0812 & 0.0035 \\
$f(\hat{\lambda})=c_0+c_1\hat{\lambda}+c_4\hat{\lambda}^2\ln(\hat{\lambda})+c_6\hat{\lambda}^2$ & $18.812(31)$ & 17.28/12 & 0.1395 & 0.0097 \\
$f(\hat{\lambda})=c_0+c_1\hat{\lambda}+c_5\hat{\lambda}^2\ln(\hat{\lambda})^2+c_6\hat{\lambda}^2$ & $18.865(24)$ & 20.29/12 & 0.0618 & 0.0021 \\
\hline
$f(\hat{\lambda})=c_0+c_1\hat{\lambda}+c_2\hat{\lambda}\ln(\hat{\lambda})+c_3\hat{\lambda}\ln(\hat{\lambda})^2+c_4\hat{\lambda}^2\ln(\hat{\lambda})$ & $18.54(18)$ & 14.40/11 & 0.2116 & 0.0150 \\
$f(\hat{\lambda})=c_0+c_1\hat{\lambda}+c_2\hat{\lambda}\ln(\hat{\lambda})+c_3\hat{\lambda}\ln(\hat{\lambda})^2+c_5\hat{\lambda}^2\ln(\hat{\lambda})^2$ & $18.49(15)$ & 14.56/11 & 0.2034 & 0.0138 \\
$f(\hat{\lambda})=c_0+c_1\hat{\lambda}+c_2\hat{\lambda}\ln(\hat{\lambda})+c_3\hat{\lambda}\ln(\hat{\lambda})^2+c_6\hat{\lambda}^2$ & $18.57(20)$ & 14.31/11 & 0.2162 & 0.0157 \\
$f(\hat{\lambda})=c_0+c_1\hat{\lambda}+c_2\hat{\lambda}\ln(\hat{\lambda})+c_4\hat{\lambda}^2\ln(\hat{\lambda})+c_5\hat{\lambda}^2\ln(\hat{\lambda})^2$ & $18.640(96)$ & 13.97/11 & 0.2347 & 0.0186 \\
$f(\hat{\lambda})=c_0+c_1\hat{\lambda}+c_2\hat{\lambda}\ln(\hat{\lambda})+c_4\hat{\lambda}^2\ln(\hat{\lambda})+c_6\hat{\lambda}^2$ & $18.61(12)$ & 14.15/11 & 0.2246 & 0.0170 \\
$f(\hat{\lambda})=c_0+c_1\hat{\lambda}+c_2\hat{\lambda}\ln(\hat{\lambda})+c_5\hat{\lambda}^2\ln(\hat{\lambda})^2+c_6\hat{\lambda}^2$ & $18.62(10)$ & 14.09/11 & 0.2282 & 0.0175 \\
$f(\hat{\lambda})=c_0+c_1\hat{\lambda}+c_3\hat{\lambda}\ln(\hat{\lambda})^2+c_4\hat{\lambda}^2\ln(\hat{\lambda})+c_5\hat{\lambda}^2\ln(\hat{\lambda})^2$ & $18.44(20)$ & 14.87/11 & 0.1886 & 0.0119 \\
$f(\hat{\lambda})=c_0+c_1\hat{\lambda}+c_3\hat{\lambda}\ln(\hat{\lambda})^2+c_4\hat{\lambda}^2\ln(\hat{\lambda})+c_6\hat{\lambda}^2$ & $18.51(18)$ & 14.49/11 & 0.2068 & 0.0143 \\
$f(\hat{\lambda})=c_0+c_1\hat{\lambda}+c_3\hat{\lambda}\ln(\hat{\lambda})^2+c_5\hat{\lambda}^2\ln(\hat{\lambda})^2+c_6\hat{\lambda}^2$ & $18.47(17)$ & 14.69/11 & 0.1970 & 0.0130 \\
$f(\hat{\lambda})=c_0+c_1\hat{\lambda}+c_4\hat{\lambda}^2\ln(\hat{\lambda})+c_5\hat{\lambda}^2\ln(\hat{\lambda})^2+c_6\hat{\lambda}^2$ & $18.685(76)$ & 13.79/11 & 0.2451 & 0.0204 \\
\hline
$f(\hat{\lambda})=c_0+c_1\hat{\lambda}+c_2\hat{\lambda}\ln(\hat{\lambda})+c_3\hat{\lambda}\ln(\hat{\lambda})^2+c_4\hat{\lambda}^2\ln(\hat{\lambda})+c_5\hat{\lambda}^2\ln(\hat{\lambda})^2$ & $19.08(54)$ & 13.31/10 & 0.2071 & 0.0095 \\
$f(\hat{\lambda})=c_0+c_1\hat{\lambda}+c_2\hat{\lambda}\ln(\hat{\lambda})+c_3\hat{\lambda}\ln(\hat{\lambda})^2+c_4\hat{\lambda}^2\ln(\hat{\lambda})+c_6\hat{\lambda}^2$ & $19.25(67)$ & 13.24/10 & 0.2107 & 0.0099 \\
$f(\hat{\lambda})=c_0+c_1\hat{\lambda}+c_2\hat{\lambda}\ln(\hat{\lambda})+c_3\hat{\lambda}\ln(\hat{\lambda})^2+c_5\hat{\lambda}^2\ln(\hat{\lambda})^2+c_6\hat{\lambda}^2$ & $19.14(59)$ & 13.28/10 & 0.2085 & 0.0097 \\
$f(\hat{\lambda})=c_0+c_1\hat{\lambda}+c_2\hat{\lambda}\ln(\hat{\lambda})+c_4\hat{\lambda}^2\ln(\hat{\lambda})+c_5\hat{\lambda}^2\ln(\hat{\lambda})^2+c_6\hat{\lambda}^2$ & $18.92(38)$ & 13.37/10 & 0.2036 & 0.0092 \\
$f(\hat{\lambda})=c_0+c_1\hat{\lambda}+c_3\hat{\lambda}\ln(\hat{\lambda})^2+c_4\hat{\lambda}^2\ln(\hat{\lambda})+c_5\hat{\lambda}^2\ln(\hat{\lambda})^2+c_6\hat{\lambda}^2$ & $19.02(51)$ & 13.33/10 & 0.2059 & 0.0094 \\
\hline
$f(\hat{\lambda})=c_0+c_1\hat{\lambda}+c_2\hat{\lambda}\ln(\hat{\lambda})+c_3\hat{\lambda}\ln(\hat{\lambda})^2+c_4\hat{\lambda}^2\ln(\hat{\lambda})+c_5\hat{\lambda}^2\ln(\hat{\lambda})^2+c_6\hat{\lambda}^2$ & $20.4(24)$ & 13.00/9 & 0.1628 & 0.0041 \\
\hline
\end{tabular}
\caption{\label{tab:lambda_max_1.8}
Results of the fits based on the linear ansatz $c_0+c_1\hat{\lambda}$ plus any of the terms with $c_2,c_3,c_4,c_5,c_6$ switched on or off. The grouping with horizontal lines indicates how many of these terms are switched on and thus $n_\mathrm{dof}$. The fitted $f_\mathrm{c}=c_0$ is given with its statistical uncertainty and the goodness of fit (normalized AIC-weights will be used below). All fits are for $\hat{\lambda}_0^\mathrm{max}=1.8$.}
\end{table}

For completeness, Tab.~\ref{tab:lambda_max_1.8} lists all $2^5=32$ fits for $\hat{\lambda}_0^\mathrm{max}=1.8$. Its columns give the ansatz, the resulting $f_\mathrm{c}$, $\chi^2/\mathrm{dof}$, $p$-value, and $w_\mathrm{AIC}$ (before normalization) as defined in Eq.~\eqref{eq:w_aic}. The linear ansatz and all single-term extensions yield unacceptable $p$-values, with the $c_4$ fit performing slightly better than the other ones. The pruning history then proceeds through $\{c_2,c_4\}$ ($p=0.1867$, $w_\mathrm{AIC} = 0.0174$) to $\{c_2,c_4,c_5\}$ ($p=0.2347$, $w_\mathrm{AIC} = 0.0186$). At this point, adding $c_3$ or $c_6$ will reduce $w_\mathrm{AIC}$ to $0.0095$ and $0.0092$, respectively; so the pruning process stops here. The fit shown in the left panel of Fig.~\ref{fig:fc_lambdamax_1.8} is the last fit in this pruning chain before degradation sets in.

After pruning, several statistically acceptable fits remain, and we lack an objective criterion to select one of them. In particular, keeping only the fit with the largest AIC weight would ignore the systematic uncertainty that comes from our ignorance of the ``true'' functional form. To combine the relevant fits and estimate the inherent systematics, we repeat the procedure used in our 2D analysis~\cite{Durr:2025hfu}, which is based on Refs.~\cite{BMW:2014pzb,Borsanyi:2020mff}. 

The fits that survive the pruning history define a set of $f_\mathrm{c}$ values that enter the CDF. At each value of $f_\mathrm{c}$, we install a step whose height is $w_\mathrm{AIC}$ (with the normalization mentioned). The resulting step function is visible as a dashed red curve in the right panel of Fig.~\ref{fig:fc_lambdamax_1.8}. At each step position, the horizontal light-red band indicates the statistical uncertainty of the associated $f_\mathrm{c}$. The median of the multi-step curve (i.e., the $x$-value where it reaches $y=0.5$) gives one possible estimate of the central value for $f_\mathrm{c}$.

The step function shown in Fig.~\ref{fig:fc_lambdamax_1.8} does not account for the statistical uncertainties of the fit results on which it is based. Therefore, we ``smooth out'' each step by replacing it with a normal CDF whose width is given by the statistical uncertainty of the pertinent $f_\mathrm{c}$. The resulting CDF is shown as a full blue curve in the right panel of Fig.~\ref{fig:fc_lambdamax_1.8}. Its median defines our central value, while the $16^\mathrm{th}$ and $84^\mathrm{th}$ percentiles define the total uncertainty. We split this uncertainty into its statistical and systematic contributions by the method described in App.~\ref{sec:app_b}. For the endpoint $\hat{\lambda}_0^\mathrm{max}=1.8$ we find $f_\mathrm{c}=18.72(11)_\mathrm{stat}(12)_\mathrm{sys}=18.72(16)_\mathrm{tot}$.

\begin{figure}[!tb]
    \includegraphics[width=0.5\linewidth]{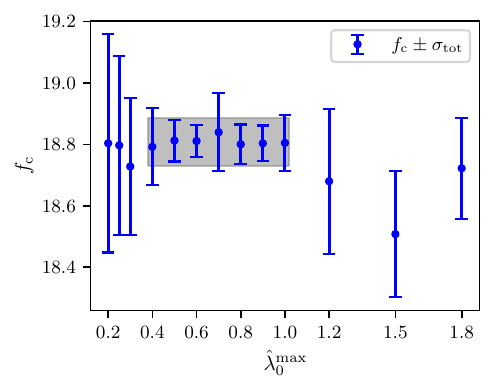}%
    \includegraphics[width=0.5\linewidth]{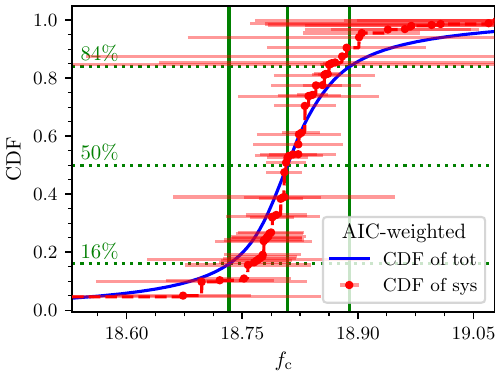}
    \caption{\label{fig:fc_final}
    Left: Values of $f_\mathrm{c}$ obtained with the procedure shown in the right panel of Fig.~\ref{fig:fc_lambdamax_1.8}, separately for $\hat{\lambda}_0^\mathrm{max} \in \{1.8,1.5,1.2,1.0, 0.9, 0.8, 0.7, 0.6, 0.5, 0.4, 0.3, 0.25,0.2\}$. Right: CDF of all statistically acceptable fits from the left panel with $\hat{\lambda}_0^\mathrm{max} \in \{1.0,0.9,0.8,0.7,0.6,0.5,0.4\}$, merged in the same manner as previously for a single $\hat{\lambda}_0^\mathrm{max}$. From this combined analysis, we obtain our final result as quoted in Eq.~\ref{eq:fc_res_final}.}
\end{figure}

The next step is to explore how things change if we exclude some of the data with large $\hat{\lambda}$ or $\hat{\lambda}_0$, i.e., drop the coarsest (infinite-volume extrapolated) data. Varying the upper end of the fit interval thus tests the robustness of the continuum extrapolation. For every choice of $\hat{\lambda}_0^\mathrm{max}\in\{1.8,1.5,1.2,1.0,0.9,0.8,0.7,0.6,0.5,0.4,0.3,0.25,0.2\}$ we repeat the pruning process in the same way as before. The resulting histories are assembled in App.~\ref{sec:app_c}. For each cut on the data, we form the CDF and determine the total uncertainty. The results are shown in the left panel of Fig.~\ref{fig:fc_final}. For $\hat{\lambda}_0^\mathrm{max}\in\{1.8,1.5,1.2\}$ the central values tend to be a bit on the low side, but this is backed by a larger total uncertainty. After the three largest $\hat{\lambda}_0$ are omitted, the central values remain remarkably stable (note that the error bars are strongly correlated). For $\hat{\lambda}_0^\mathrm{max}\in\{0.4,0.3,0.25,0.2\}$ the number of degrees of freedom decreases, and the total uncertainty increases distinctively. The uncertainty is minimal at $\hat{\lambda}_0^\mathrm{max}=0.6$.

An aggressive choice would be to take the value of $f_\mathrm{c}$ at $\hat{\lambda}_0^\mathrm{max}=0.6$ as our final estimate. To be more conservative, we merge all fits retained by the pruning histories for $\hat{\lambda}_0^\mathrm{max}\in\{1.0,0.9,0.8,0.7,0.6,0.5,0.4\}$. This range covers the entire stable region. The combined CDF is shown in the right panel of Fig.~\ref{fig:fc_final}. Its median and central $68\%$-percentiles give our final result
\begin{equation}
    f_\mathrm{c} = 18.808(64)_{\mathrm{stat}}(45)_{\mathrm{sys}} = 18.808(78)_{\mathrm{tot}}
    \label{eq:fc_res_final}
\end{equation}
where we have taken the liberty to symmetrize the up/down-uncertainties.
This result is shown as the gray horizontal band in the left panel of Fig.~\ref{fig:fc_final}.


\section{Continuum limit of \ensuremath{\Delta\langle\phi^2\rangle_\mathrm{c}/\hat{\lambda}}\label{sec:cont_dphisq}}

\begin{figure}[!tb]
    \includegraphics[width=0.5\linewidth]{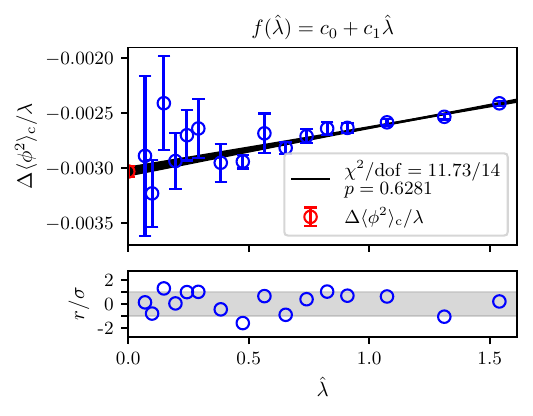}%
    \includegraphics[width=0.5\linewidth]{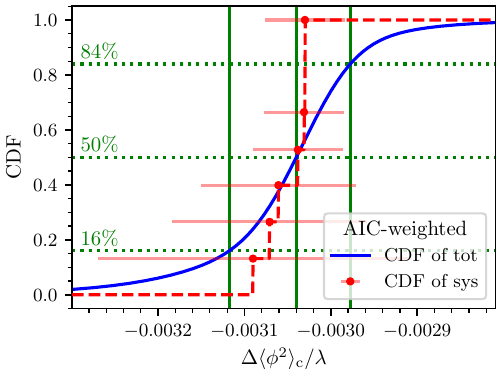}
    \caption{\label{fig:dphisq_lambdamax_1.8}Left: Example of a continuum extrapolation of $\Delta\langle\phi^2\rangle_\mathrm{c}/\hat\lambda$ with a $\hat\lambda^\mathrm{max}$ that corresponds to $\hat{\lambda}_0^{\max}=1.8$, using the linear ansatz. Right: CDF of the six fits at $\hat{\lambda}_0^{\max}=1.8$ (linear ansatz and its five single-term extensions $\{c_2\},\{c_3\},\{c_4\},\{c_5\},\{c_6\}$), weighted by their AIC scores. Unlike for $f_\mathrm{c}$ in Fig.~\ref{fig:fc_lambdamax_1.8}, no further combinations of two or more additional terms are included here, since the statistical uncertainty of $\Delta\langle\phi^2\rangle_\mathrm{c}/\hat{\lambda}$ proliferates at small $\hat\lambda$, so that such fits would predominantly fit noise rather than resolve genuine curvature.}
\end{figure}

\begin{table}[!tb]
\centering
\begin{tabular}{@{\hskip 5pt}l@{\hskip 5pt}c@{\hskip 5pt}c@{\hskip 5pt}c@{\hskip 5pt}c@{\hskip 5pt}|l@{\hskip 5pt}c@{\hskip 5pt}c@{\hskip 5pt}c@{\hskip 5pt}c@{\hskip 5pt}}
\hline
fit ansatz & $\Delta\langle\phi^2\rangle_\mathrm{c}/\lambda$ & $\chi^2/\mathrm{dof}$ & $p$ & $w_\mathrm{AIC}$ & fit ansatz & $\Delta\langle\phi^2\rangle_\mathrm{c}/\lambda$ & $\chi^2/\mathrm{dof}$ & $p$ & $w_\mathrm{AIC}$ \\
\hline
linear & $-0.003030(46)$ & 11.73/14 & 0.6281 & 1.1451 & linear & $-0.003017(55)$ & 11.57/13 & 0.5630 & 0.7510 \\
\hline
$\{c_{2}\}$ & $-0.00309(18)$ & 11.60/13 & 0.5605 & 0.4488 & $\{c_{2}\}$ & $-0.00320(21)$ & 10.78/12 & 0.5476 & 0.4101 \\
$\{c_{3}\}$ & $-0.003031(46)$ & 11.52/13 & 0.5674 & 0.4678 & $\{c_{3}\}$ & $-0.002961(71)$ & 9.727/12 & 0.6399 & 0.6951 \\
$\{c_{4}\}$ & $-0.003061(89)$ & 11.57/13 & 0.5629 & 0.4554 & $\{c_{4}\}$ & $-0.00314(11)$ & 9.762/12 & 0.6368 & 0.6830 \\
$\{c_{5}\}$ & $-0.003038(52)$ & 11.64/13 & 0.5570 & 0.4396 & $\{c_{5}\}$ & $-0.003008(56)$ & 8.287/12 & 0.7623 & 1.4284 \\
$\{c_{6}\}$ & $-0.00307(11)$ & 11.57/13 & 0.5632 & 0.4562 & $\{c_{6}\}$ & $-0.00317(14)$ & 10.19/12 & 0.5992 & 0.5512 \\
\hline
\end{tabular}
\caption{\label{tab:lambda_max_1.8_1.5_phisq}
Results of the linear ansatz and its five single-term extensions $\{c_2\},\{c_3\},\{c_4\},\{c_5\},\{c_6\}$ for $\Delta\langle\phi^2\rangle_\mathrm{c}/\lambda$ with $\hat{\lambda}_0^\mathrm{max}=1.8$ (left) and $\hat{\lambda}_0^\mathrm{max}=1.5$ (right). No pruning is applied and no combinations of two or more additional terms are considered, since the larger statistical uncertainties in the data at small $\hat\lambda$ make such extended ans\"atze prone to fitting noise rather than genuine curvature.}
\end{table}

\begin{figure}[!tb]
    \includegraphics[width=0.5\linewidth]{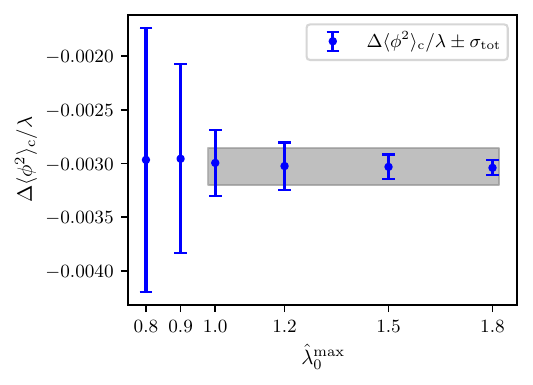}%
    \includegraphics[width=0.5\linewidth]{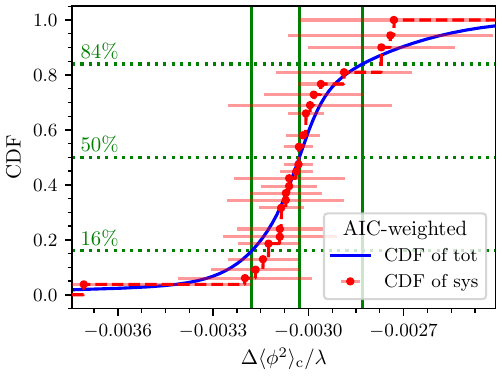}
    \caption{\label{fig:dphisq_final}
        Left: Values of $\Delta\langle\phi^2\rangle_\mathrm{c}/\lambda$ obtained with the procedure of Fig.~\ref{fig:dphisq_lambdamax_1.8}, combining the linear ansatz with its five single-term extensions, separately for $\hat{\lambda}_0^\mathrm{max} \in \{1.8,1.5,1.2,1.0,0.9,0.8\}$. Right: CDF obtained by merging all results from the left panel except $\hat{\lambda}_0^\mathrm{max}\in\{0.8,0.9\}$, in the same manner as for a single $\hat\lambda_0^{\max}$ in Fig.~\ref{fig:dphisq_lambdamax_1.8}. From this combined analysis we obtain our final result as quoted in Eq.~\eqref{eq:dphisq_final}.}
\end{figure}

To take the continuum limit of $\Delta\langle\phi^2\rangle_\mathrm{c}/\hat{\lambda}$ we proceed in the same manner as for $f_\mathrm{c}$. Here, the uncertainties increase even more rapidly towards small $\hat{\lambda}$. Again we use the function base of Eq.~\ref{eq:fc_cont_fit}, but this time we restrict the survey to the linear ansatz and the five single-term extensions $\{c_2\},\ldots,\{c_6\}$. Trial fits with additional terms give poorly determined coefficients $c_i$ and do not resolve any further curvature. Hence, no iterative pruning is needed here. For every $\hat{\lambda}_0^\mathrm{max}$, the six fits are combined with the same AIC-weighted CDF procedure as in Sec.~\ref{sec:cont_fc}.

We first consider $\hat{\lambda}_0^\mathrm{max}=1.8$. The linear fit is shown in the left panel of Fig.~\ref{fig:dphisq_lambdamax_1.8}, and it yields $\chi^2/\mathrm{dof}=11.73/14$ and $p=0.6281$. The linear ansatz and the five extensions are listed in the left half of Tab.~\ref{tab:lambda_max_1.8_1.5_phisq}. All six fits yield an acceptable description of the data and none of the additional terms is strongly preferred. Therefore, we include all six fits in the CDF shown in the right panel of Fig.~\ref{fig:dphisq_lambdamax_1.8}.

We repeat this analysis for every $\hat{\lambda}_0^\mathrm{max}\in\{1.5,1.2,1.0,0.9,0.8\}$. The results are collected in Tabs.~\ref{tab:lambda_max_1.2_1.0_phisq} and \ref{tab:lambda_max_0.9_0.8_phisq} in App.~\ref{sec:app_c}. We do not continue toward smaller $\hat{\lambda}_0^\mathrm{max}$, since the uncertainties of the remaining data increase rapidly below $\hat{\lambda}^\mathrm{max}\simeq0.6$.

The outcome of the cutoff scan is shown in the left panel of Fig.~\ref{fig:dphisq_final}. The central value is stable over the full range (again, the error bars are strongly correlated). The smallest total uncertainty is obtained at $\hat{\lambda}_0^\mathrm{max}=1.8$. This uncertainty increases monotonically for decreasing $\hat{\lambda}_0^\mathrm{max}$, and increases drastically for $\hat{\lambda}_0^\mathrm{max}\leq0.9$.
The final analysis includes the four choices $\hat{\lambda}_0^\mathrm{max}\in\{1.0,1.2,1.5,1.8\}$. The combined AIC-weighted CDF of all 24 fits is shown in the right panel of Fig.~\ref{fig:dphisq_final}. Its median and central $68\%$ interval give the final result
\begin{equation}
    \Delta\langle\phi^2\rangle_\mathrm{c}/\lambda = -0.00303(12)_{\mathrm{stat}}(12)_{\mathrm{sys}} = -0.00303(17)_{\mathrm{tot}} \;.
    \label{eq:dphisq_final} 
\end{equation}


\section{Conclusion\label{sec:conc}}

The goal of this paper has been to determine the critical coupling $f_\mathrm{c}$ as defined in (\ref{def_fc}) and the quantity $\Delta\langle\phi^2\rangle_\mathrm{c}/\lambda$ as defined in (\ref{eq:dphisq_lattice_limit}) for the scalar $\phi^4$ theory in 3D, in close analogy to our 2D paper~\cite{Durr:2025hfu}.
Our final results for these quantities have been presented in Eq.~(\ref{eq:fc_res_final}) and Eq.~(\ref{eq:dphisq_final}), respectively.
We have spent a significant amount of effort to remove, by means of sophisticated extrapolations, the effect of finite box volumes and finite lattice spacings.
The result is a universal value of $f_\mathrm{c}$ that is free of any remains of the lattice formulation that was used in the intermediate steps of the computation, and a similar statement holds for $\Delta\langle\phi^2\rangle_\mathrm{c}/\lambda$.

\begin{table}[!tb]
    \centering
    \begin{tabular}{llcc}
    \hline
    Reference & Method & $f_\mathrm{c}$ & $\Delta\langle\phi^2\rangle_\mathrm{c}/\lambda$ \\
    \hline
    Sun (2003)~\cite{Sun:2002cc} & Lattice Monte Carlo & $18.216(57)$ & $-0.00296(25)$\\
    Kastening (2004)~\cite{Kastening:2004wu,Kastening:2003iu} & 7-loop variational PT & $18.30(18)$ & $-0.00291(27)$ \\
    Blaizot, M\'endez-Galain, Wschebor (2006)~\cite{Blaizot:2005wd} & Approximate RG equations & x & $-0.00288(73)$\\
    Pelissetto and Vicari (2015)~\cite{Pelissetto:2015yha} & Borel-resummed PT & $18.461(74)$ & x \\
    Sberveglieri, Serone, Spada (2021)~\cite{Sberveglieri:2020eko} & Perturbation theory & $18.7(10)$ & x \\ 
    Yukalov and Yukalova (2022)~\cite{Yukalov:2022gbr} & Self-similar approximants & x & $-0.00288(14)$ \\
    this work (2026)  & Lattice Monte Carlo & $18.808(78)$ & $-0.00303(17)$ \\
    \hline
    \end{tabular}
    \caption{\label{tab:literature}Comparison with the literature on $f_\mathrm{c}$ and $\Delta\langle\phi^2\rangle_\mathrm{c}/\lambda$, always quoting the total uncertainty.}
\end{table}

\begin{figure}[!tb]
    \centering
    \includegraphics[width=0.9\linewidth]{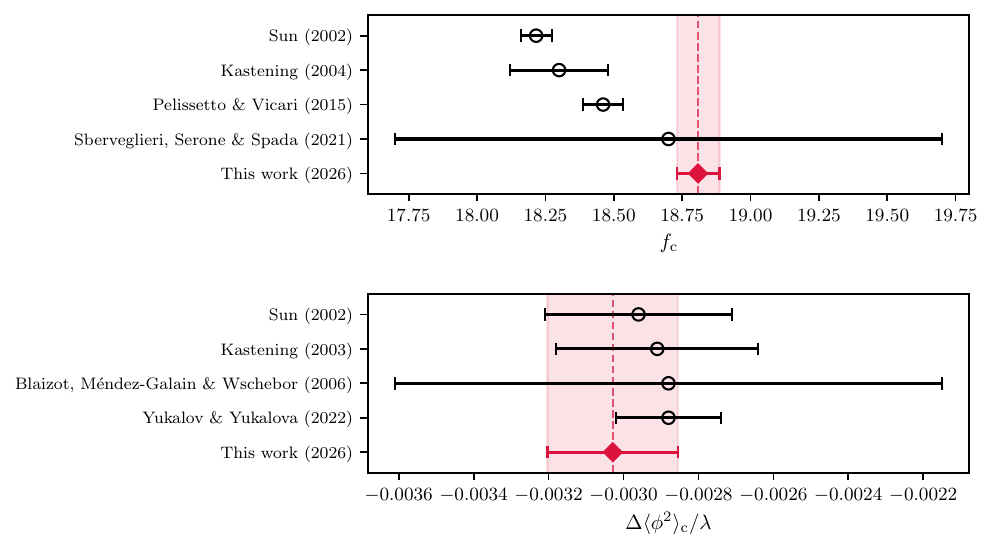}\\
    \caption{Comparison with literature values on $f_\mathrm{c}$ (top) and $\Delta\langle\phi^2\rangle_\mathrm{c}/\lambda$ (bottom). See Tab.~\ref{tab:literature} for details.}
    \label{fig:literature_combined}
\end{figure}

Unfortunately, we are only aware of a handful of papers that calculate $f_\mathrm{c}$ and/or $\Delta\langle\phi^2\rangle_\mathrm{c}/\lambda$ in 3D.
We list them in Tab.~\ref{tab:literature}, along with a keyword to characterize the (main) computational approach.
The same information is presented in graphical form in Fig.~\ref{fig:literature_combined}.
The graphical version makes it clear that $f_\mathrm{c}$ has been drifting upward since its initial determination by Sun~\cite{Sun:2002cc}.
On the other hand, the computations of $\Delta\langle\phi^2\rangle_\mathrm{c}/\lambda$ have been remarkably stable over time, both in terms of the central value and the size of the uncertainty.

We used the same sophisticated machinery for assessing statistical and systematic errors as in our 2D analysis~\cite{Durr:2025hfu}.
In retrospect, it is not entirely clear whether this is really needed here, since the continuum extrapolation curve in 3D (plotted in Fig.~\ref{fig:fc_final}) shows much less curvature than its counterpart in 2D.
In our judgment the slight tension with the result by Sun~\cite{Sun:2002cc} is unlikely to originate from our more sophisticated fitting ans\"atze, but rather from the extra computer power that we had at our disposal to explore significantly larger $N_x$-values and thus avoid any inaccurate assessment in the infinite-volume extrapolations (which tend to mimic curvature at small values of $\hat{\lambda}_0$ or $\hat{\lambda}$). Obviously, our ansatz Eq.~\eqref{eq:fc_cont_fit} recycles an ansatz that was successful in 2D, but any analytical guidance on which terms are really needed in 3D would be very useful.

An interesting extension of this work would be to consider $N$-component $\phi^4$ theories in 3D; for $N=2$ this was done in Ref.~\cite{Arnold:2001ir}, and for $N=4$ in Ref.~\cite{Sun:2002cc}.
In particular, the case $N=2$ is of interest to ``real life physics'', since it relates to Bose-Einstein condensation (BEC).
The result $\Delta\langle\phi^2\rangle_\mathrm{c}/\lambda=-0.00719(10)$ in Ref.~\cite{Arnold:2001ir} settled the long-standing debate about the numerical value of the BEC temperature shift coefficient ($c\simeq1.32$).

From a lattice perspective, an open issue is whether there is another discretization of the Laplacian in (\ref{def_cont}) that facilitates the continuum extrapolation (e.g.\ by a smaller slope and/or less curvature in Fig.~\ref{fig:fc_lambdamax_1.8}, while the extrapolated value $f_\mathrm{c}$ would remain unaffected, of course).
In this context, we recall that Sun in Ref.~\cite{Sun:2002cc} used a 13-point stencil, while we use the standard 7-point stencil in 3D.
From comparing Fig.~7a in that paper%
\footnote{To the best of our knowledge, there is a labeling mistake in that figure. We think that the (black) circles refer to the ``O(4) $Lu=144$ data'', while the (red) diamonds refer to the ``O(1) $Lu=768$ data''. With this correction, we find complete agreement with the raw data in various tables of that paper.}
to the left panel in our Fig.~\ref{fig:fc_lambdamax_1.8}, one does not get the impression that the 13-point stencil would entail any noticeable advantage, but a more carefully chosen lattice Laplacian might possibly change that.


\appendix


\section{Finite-volume scaling ans\"atze\label{sec:app_a}}

The finite-size scaling arguments used in this paper are standard and may be found in many textbooks and reviews, see, e.g., Ref.~\cite{Pelissetto:2000ek}. We review the steps needed to understand our fit ans\"atze for completeness.

For the scaling behavior of the pseudocritical mass parameter $\hat{\mu}_0^2$ we use the expansion of RG-invariant quantities given in Ref.~\cite{Campostrini:2006ms}, while for the derivation for $\langle\phi^2\rangle_{\mathrm{c}}$ we follow Ref.~\cite{Arnold:2001ir}.
Throughout this appendix, $\hat{\lambda}_0$ is kept fixed.
We define the dimensionless distance from the infinite-volume critical point by
\begin{equation}
t=\hat{\mu}_0^2-\hat{\mu}_{0\mathrm{c}}^2(\hat{\lambda}_0) \;.
\label{eq:fss_t}
\end{equation}

Let us first consider the pseudocritical mass. The Binder cumulant is an RG-invariant quantity. Its finite-size scaling expansion can therefore be written as (see Eq.~(15) in Ref.~\cite{Campostrini:2006ms})
\begin{equation}
U_4(t,N_x)=\mathcal{U}_0(u_tN_x^{y_t})+\sum_{n\geq1}\mathcal{U}_{3,n}(u_tN_x^{y_t})u_3^nN_x^{ny_3}+\sum_{j\geq4}\mathcal{U}_j(u_tN_x^{y_t})u_jN_x^{y_j}+\cdots,\qquad y_t=\frac{1}{\nu} \;.
\label{eq:binder_scaling_form}
\end{equation}
Here $u_3$ is the leading irrelevant scaling field, with exponent $y_3=-\omega$, while every $u_j$ with $j\geq4$ denotes further irrelevant fields with exponents $y_j<0$. The first sum contains successive powers of the leading irrelevant field. The functions $\mathcal{U}_0(x),\;\mathcal{U}_j(x)$ and $\mathcal{U}_{3,n}(x)$ are smooth, finite for $x\to0$. The scaling fields $u_t(t,\hat{\lambda}_0)$ and $u_i(t,\hat{\lambda}_0)$ are analytic functions of the model parameters, with $u_t\sim t$ and the irrelevant fields $u_i$ generically finite at $t=0$.

At the infinite-volume critical point, $u_tN_x^{y_t}\sim tN_x^{y_t} \to 0$ and all finite-size corrections vanish, so that $U_4^*\equiv\mathcal{U}_0(0)$. For every $N_x$, our scan in $\hat{\mu}_0^2$ determines the pseudocritical mass through the relation
\begin{equation}
U_4(t_{\mathrm{c}},N_x)=U_4^* \;,\qquad t_{\mathrm{c}}=\hat{\mu}_{0\mathrm{c}}^2(\hat{\lambda}_0,N_x)-\hat{\mu}_{0\mathrm{c}}^2(\hat{\lambda}_0) \;.
\label{eq:binder_condition}
\end{equation}
At fixed $\hat{\lambda}_0$, the irrelevant scaling fields are constant and their powers can be absorbed into the coefficients. Expanding the scaling functions around $x=0$, imposing Eq.~\eqref{eq:binder_condition}, and solving order by order in the leading correction field gives
\begin{equation}
\hat{\mu}_{0\mathrm{c}}^2(\hat{\lambda}_0,N_x)=\hat{\mu}_{0\mathrm{c}}^2(\hat{\lambda}_0)+c_1N_x^{-y_t-\omega}+c_2N_x^{-y_t-2\omega}+c_3N_x^{-y_t-3\omega}+\cdots \;.
\label{eq:mu0sq_fss_derived}
\end{equation}
The displayed terms are the first three powers of the leading irrelevant field. They should not be understood as the first three terms in the complete asymptotic expansion, since corrections from further irrelevant fields and from the analytic background may occur between them.

We now turn to $\langle\phi^2\rangle_{\mathrm{c}}$. Following Eq.~(4.3) of Ref.~\cite{Arnold:2001ir}, we choose the rescaling factor as $b=N_x$ and retain only the leading irrelevant field. The dimensionless free-energy density per lattice site then takes the form
\begin{equation}
f(t,N_x)=f_{\mathrm{reg}}(t)+N_x^{-d}\mathcal{F}(x,z) \;,\qquad x=N_x^{y_t}t \;,\qquad z=N_x^{-\omega}
\label{eq:free_energy_fss}
\end{equation}
where $f_{\mathrm{reg}}$ and $\mathcal{F}$ denote the regular and singular contributions, respectively. Expanding both contributions about the critical point, and collecting equal powers of $t$ gives
\begin{align}
f(t,N_x)&=A_0+B_0N_x^{-d}+C_0N_x^{-d-\omega}+\cdots\nonumber\\
&+t\left(A_1+B_1N_x^{-d+y_t}+C_1N_x^{-d+y_t-\omega}+\cdots\right)\nonumber\\
&+t^2\left(A_2+B_2N_x^{-d+2y_t}+\cdots\right)+\cdots \;.
\label{eq:free_energy_expansion}
\end{align}
Since $t$ enters the action through the mass term, it follows that
\begin{equation}
\frac{\partial f}{\partial t}=\frac{1}{2N_x^d}\left\langle\sum_{y\in\Lambda}\phi(y)^2\right\rangle=\frac{1}{2}\langle\phi^2\rangle .
\label{eq:free_energy_derivative}
\end{equation}
Differentiating Eq.~\eqref{eq:free_energy_expansion} and inserting the finite-size expansion of $t_{\mathrm{c}}$ from Eq.~\eqref{eq:mu0sq_fss_derived} yields
\begin{equation}
\langle\phi^2\rangle_{\mathrm{c}}=A+BN_x^{-d+y_t}+\sum_{n\geq1}\left(C_nN_x^{-d+y_t-n\omega}+D_nN_x^{-y_t-n\omega}\right)+\cdots .
\label{eq:phisq_fss_full}
\end{equation}
The terms proportional to $C_n$ arise from the singular part of the free energy, while those proportional to $D_n$ arise from its regular part evaluated at the pseudocritical mass. For the 3D Ising exponents, the two powers for a given $n$ are numerically close to each other. This is why we take the liberty to replace them by a single term with their mean decay exponent, $d/2+n\omega$, and our functional form thus reads
\begin{equation}
\langle\phi^2\rangle_{\mathrm{c}}(N_x)=A+BN_x^{-d+y_t}+C^\prime N_x^{-d/2-\omega}+D^\prime N_x^{-d/2-2\omega} \;.
\label{eq:phisq_fss_effective_4_param}
\end{equation}
For $d=3$ and $y_t=1/\nu$, this is the four-parameter ansatz used in Eq.~\eqref{eq:phisq_fit_ansatz}. The versions with $n_{\mathrm{par}}=3$ and $n_{\mathrm{par}}=2$ follow by setting $D^\prime=0$ and, in addition, $C^\prime=0$, respectively. 


\section{Assessment of statistical and systematic uncertainties\label{sec:app_b}}

This procedure follows the analysis introduced in Ref.~\cite{BMW:2014pzb} with the refinements of Ref.~\cite{Borsanyi:2020mff}.
We summarize it here for completeness and to provide an easily accessible reference for the method used.

To estimate the statistical and systematic components of the total uncertainty, we treat the set of AIC-weighted fits as a Gaussian mixture.
Each fit $i$ contributes a Gaussian distribution with central value $f_{\mathrm{c},i}$ and statistical width $\sigma_i$, weighted by $w_i$.
The resulting probability density is
\begin{equation}
    \sum_i w_i\, N(y; f_{\mathrm{c},i}, \sigma_i)
\end{equation}
which corresponds to the cumulative distribution function (CDF)
\begin{equation}
    P(y;\kappa)=\int_{-\infty}^{y}\mathrm{d}y^\prime~\sum_i w_i\,N(y^\prime; f_{\mathrm{c},i},\sigma_i\sqrt{\kappa})\;,
\end{equation}
where the parameter $\kappa$ rescales the statistical widths without modifying the spread of the central values.
The total uncertainty is defined from the $16^{\mathrm{th}}$ and $84^{\mathrm{th}}$ percentiles of the CDF evaluated at $\kappa=1$,
\begin{equation}
    \sigma^2_{\mathrm{tot}} \equiv \left[\frac{1}{2}(y_{84}-y_{16})\right]^2, \qquad\text{with}\qquad P(y_{16};1)=0.16,\quad P(y_{84};1)=0.84\;.
\end{equation}
This uncertainty is decomposed into statistical and systematic contributions with
\begin{equation}
    \sigma_{\mathrm{stat}}^2 + \sigma_{\mathrm{sys}}^2 \equiv \sigma_{\mathrm{tot}}^2 \;.
\end{equation}
To isolate the statistical component, we artificially rescale the statistical widths by a
factor $\sqrt{\kappa}$ and extract the corresponding percentile interval
\begin{equation}
    \kappa\sigma_{\mathrm{stat}}^2 + \sigma_{\mathrm{sys}}^2 \equiv \left[\frac{1}{2}(\tilde{y}_{84}-\tilde{y}_{16})\right]^2 \qquad\text{with}\qquad P(\tilde{y}_{16};\kappa)=0.16,\quad P(\tilde{y}_{84};\kappa)=0.84 \;.
\end{equation}
Taking into account the difference between the expression with arbitrary $\kappa$ and the one evaluated at $\kappa=1$ yields
\begin{equation}
    \left[\frac{1}{2}(\tilde{y}_{84}-\tilde{y}_{16})\right]^2-\left[\frac{1}{2}(y_{84}-y_{16})\right]^2 = (\kappa-1)\sigma_\mathrm{stat}^2 \;.
\end{equation}
The latter equality provides a direct relation between the change in percentile width under the rescaling and the underlying statistical variance (for our analysis we choose $\kappa=2$).


\section{Pruned fits \label{sec:app_c}}

Tables~\ref{tab:lambda_max_1.5_1.2}, \ref{tab:lambda_max_1.0_0.9}, \ref{tab:lambda_max_0.8_0.7}, \ref{tab:lambda_max_0.6_0.5}, \ref{tab:lambda_max_0.4_0.3} and \ref{tab:lambda_max_0.25_0.2} summarize the pruned fits entering the continuum extrapolation of $f_\mathrm{c}$, while tables~\ref{tab:lambda_max_1.2_1.0_phisq} and \ref{tab:lambda_max_0.9_0.8_phisq} summarize the fits entering the continuum extrapolation of $\Delta\langle\phi^2\rangle_\mathrm{c}/\lambda$.

\begin{table}[!h]
\centering
\begin{tabular}{@{\hskip 5pt}l@{\hskip 5pt}c@{\hskip 5pt}c@{\hskip 5pt}c@{\hskip 5pt}c@{\hskip 5pt}|@{\hskip 5pt}l@{\hskip 5pt}c@{\hskip 5pt}c@{\hskip 5pt}c@{\hskip 5pt}c@{\hskip 5pt}}
\hline
fit ansatz & $f_\mathrm{c}$ & $\chi^2/\mathrm{dof}$ & $p$ & $w_\mathrm{AIC}$ & fit ansatz & $f_\mathrm{c}$ & $\chi^2/\mathrm{dof}$ & $p$ & $w_\mathrm{AIC}$ \\
\hline
linear & $18.7502(58)$ & 282.2/13 & 0.0000 & 0.0000 & linear & $18.8098(69)$ & 59.19/12 & 0.0000 & 0.0000 \\
\hline
$\{c_{2}\}$ & $19.081(24)$ & 76.02/12 & 0.0000 & 0.0000 & $\{c_{2}\}$ & $18.955(31)$ & 35.75/11 & 0.0002 & 0.0000 \\
$\{c_{3}\}$ & $18.6842(73)$ & 25.60/12 & 0.0122 & 0.0002 & $\{c_{3}\}$ & $18.716(17)$ & 21.35/11 & 0.0299 & 0.0013 \\
$\{c_{4}\}$ & $18.904(11)$ & 27.03/12 & 0.0077 & 0.0001 & $\{c_{4}\}$ & $18.887(14)$ & 23.87/11 & 0.0133 & 0.0004 \\
$\{c_{5}\}$ & $18.7401(59)$ & 30.37/12 & 0.0025 & 0.0000 & $\{c_{5}\}$ & $18.684(23)$ & 23.5/11 & 0.0150 & 0.0004 \\
$\{c_{6}\}$ & $18.959(14)$ & 41.84/12 & 0.0000 & 0.0000 & $\{c_{6}\}$ & $18.911(19)$ & 28.72/11 & 0.0025 & 0.0000 \\
\hline
$\{c_{2},c_{3}\}$ & $18.389(93)$ & 15.19/11 & 0.1739 & 0.0166 & $\{c_{2},c_{3}\}$ & $18.41(13)$ & 15.14/10 & 0.1272 & 0.0104 \\
$\{c_{3},c_{4}\}$ & $18.64(23)$ & 25.55/11 & 0.0076 & 0.0001 & $\{c_{3},c_{4}\}$ & $18.35(26)$ & 19.15/10 & 0.0384 & 0.0014 \\
$\{c_{3},c_{5}\}$ & $18.708(10)$ & 15.59/11 & 0.1571 & 0.0136 & $\{c_{3},c_{5}\}$ & $18.670(23)$ & 12.02/10 & 0.2837 & 0.0493 \\
$\{c_{3},c_{6}\}$ & $18.28(14)$ & 16.81/11 & 0.1137 & 0.0074 & $\{c_{3},c_{6}\}$ & $18.35(17)$ & 16.00/10 & 0.0997 & 0.0068 \\
\hline
$\{c_{2},c_{3},c_{4}\}$ & $18.58(23)$ & 14.31/10 & 0.1591 & 0.0095 & $\{c_{2},c_{3},c_{5}\}$ & $18.84(25)$ & 11.59/9 & 0.2376 & 0.0225 \\
$\{c_{2},c_{3},c_{5}\}$ & $18.52(19)$ & 14.48/10 & 0.1520 & 0.0087 & $\{c_{3},c_{4},c_{5}\}$ & $18.88(30)$ & 11.56/9 & 0.2390 & 0.0228 \\
$\{c_{2},c_{3},c_{6}\}$ & $18.62(26)$ & 14.22/10 & 0.1630 & 0.0099 & $\{c_{3},c_{5},c_{6}\}$ & $18.86(28)$ & 11.58/9 & 0.2383 & 0.0226 \\
\hline
\end{tabular}
\caption{\label{tab:lambda_max_1.5_1.2}Same as Tab. \ref{tab:lambda_max_1.8}, but for $\hat{\lambda}_0^\mathrm{max}=1.5$ (left) and $\hat{\lambda}_0^\mathrm{max}=1.2$ (right) and pruned. The pruning acts whenever the number of degrees of freedom (``dof'') is reduced. The AIC-weights will be normalized later.}
\end{table}

\begin{table}[!htbp]
\centering
\begin{minipage}[t]{0.49\textwidth}
\centering
\begin{tabular}{@{\hskip 5pt}l@{\hskip 5pt}c@{\hskip 5pt}c@{\hskip 5pt}c@{\hskip 5pt}c@{\hskip 5pt}}
\hline
fit ansatz & $f_\mathrm{c}$ & $\chi^2/\mathrm{dof}$ & $p$ & $w_\mathrm{AIC}$ \\
\hline
linear & $18.8574(95)$ & 17.46/11 & 0.0951 & 0.0146 \\
\hline
$\{c_{2}\}$ & $18.837(41)$ & 17.19/10 & 0.0702 & 0.0061 \\
$\{c_{3}\}$ & $18.823(49)$ & 17.03/10 & 0.0738 & 0.0066 \\
$\{c_{4}\}$ & $18.861(18)$ & 17.41/10 & 0.0658 & 0.0055 \\
$\{c_{5}\}$ & $18.787(43)$ & 14.29/10 & 0.1600 & 0.0261 \\
$\{c_{6}\}$ & $18.854(26)$ & 17.44/10 & 0.0652 & 0.0054 \\
\hline
$\{c_{2},c_{5}\}$ & $18.810(44)$ & 11.25/9 & 0.2587 & 0.0438 \\
$\{c_{3},c_{5}\}$ & $18.673(88)$ & 12.02/9 & 0.2122 & 0.0299 \\
$\{c_{4},c_{5}\}$ & $18.768(46)$ & 11.55/9 & 0.2396 & 0.0377 \\
$\{c_{5},c_{6}\}$ & $18.776(44)$ & 11.47/9 & 0.2447 & 0.0393 \\
\hline
$\{c_{2},c_{3},c_{5}\}$ & $19.07(31)$ & 10.73/8 & 0.2177 & 0.0210 \\
$\{c_{2},c_{4},c_{5}\}$ & $19.00(21)$ & 10.66/8 & 0.2216 & 0.0217 \\
$\{c_{2},c_{5},c_{6}\}$ & $19.01(23)$ & 10.68/8 & 0.2208 & 0.0215 \\
\hline
\end{tabular}
\end{minipage}%
\begin{minipage}[t]{0.49\textwidth}
\centering
\begin{tabular}{@{\hskip 5pt}l@{\hskip 5pt}c@{\hskip 5pt}c@{\hskip 5pt}c@{\hskip 5pt}c@{\hskip 5pt}}
\hline
fit ansatz & $f_\mathrm{c}$ & $\chi^2/\mathrm{dof}$ & $p$ & $w_\mathrm{AIC}$ \\
\hline
linear & $18.866(12)$ & 15.73/10 & 0.1077 & 0.0210 \\
\hline
$\{c_{2}\}$ & $18.776(50)$ & 12.24/9 & 0.1999 & 0.0441 \\
$\{c_{3}\}$ & $18.961(96)$ & 14.98/9 & 0.0914 & 0.0112 \\
$\{c_{4}\}$ & $18.841(20)$ & 13.50/9 & 0.1414 & 0.0236 \\
$\{c_{5}\}$ & $18.785(43)$ & 11.40/9 & 0.2492 & 0.0672 \\
$\{c_{6}\}$ & $18.816(31)$ & 12.74/9 & 0.1745 & 0.0343 \\
\hline
$\{c_{2},c_{5}\}$ & $18.802(50)$ & 11.20/8 & 0.1907 & 0.0273 \\
$\{c_{3},c_{5}\}$ & $18.77(14)$ & 11.39/8 & 0.1805 & 0.0248 \\
$\{c_{4},c_{5}\}$ & $18.778(52)$ & 11.32/8 & 0.1842 & 0.0257 \\
$\{c_{5},c_{6}\}$ & $18.781(47)$ & 11.3/8 & 0.1854 & 0.0260 \\
\hline
\end{tabular}
\end{minipage}
\caption{\label{tab:lambda_max_1.0_0.9}
Same as Tab.~\ref{tab:lambda_max_1.5_1.2}, but for $\hat{\lambda}_0^\mathrm{max}=1.0$ (left) and $\hat{\lambda}_0^\mathrm{max}=0.9$ (right).}
\end{table}

\begin{table}[!h]
\centering
\begin{tabular}{@{\hskip 5pt}l@{\hskip 5pt}c@{\hskip 5pt}c@{\hskip 5pt}c@{\hskip 5pt}c@{\hskip 5pt}|@{\hskip 5pt}l@{\hskip 5pt}c@{\hskip 5pt}c@{\hskip 5pt}c@{\hskip 5pt}c@{\hskip 5pt}}
\hline
fit ansatz & $f_\mathrm{c}$ & $\chi^2/\mathrm{dof}$ & $p$ & $w_\mathrm{AIC}$ & fit ansatz & $f_\mathrm{c}$ & $\chi^2/\mathrm{dof}$ & $p$ & $w_\mathrm{AIC}$ \\
\hline
linear & $18.864(13)$ & 15.63/9 & 0.0750 & 0.0134 & linear & $18.858(17)$ & 15.28/8 & 0.0540 & 0.0097 \\
\hline
$\{c_{2}\}$ & $18.752(61)$ & 11.39/8 & 0.1808 & 0.0411 & $\{c_{2}\}$ & $18.721(70)$ & 10.45/7 & 0.1645 & 0.0398 \\
$\{c_{3}\}$ & $19.08(20)$ & 14.48/8 & 0.0700 & 0.0087 & $\{c_{3}\}$ & $19.20(31)$ & 14.3/7 & 0.0461 & 0.0058 \\
$\{c_{4}\}$ & $18.826(24)$ & 11.72/8 & 0.1642 & 0.0348 & $\{c_{4}\}$ & $18.809(26)$ & 8.3/7 & 0.3069 & 0.1165 \\
$\{c_{5}\}$ & $18.782(45)$ & 11.23/8 & 0.1890 & 0.0444 & $\{c_{5}\}$ & $18.773(47)$ & 10.77/7 & 0.1490 & 0.0339 \\
$\{c_{6}\}$ & $18.794(39)$ & 11.38/8 & 0.1809 & 0.0411 & $\{c_{6}\}$ & $18.765(46)$ & 9.577/7 & 0.2138 & 0.0615 \\
\hline
$\{c_{2},c_{5}\}$ & $18.80(14)$ & 11.20/7 & 0.1302 & 0.0166 & $\{c_{2},c_{4}\}$ & $18.97(14)$ & 7.136/6 & 0.3084 & 0.0767 \\
$\{c_{3},c_{5}\}$ & $18.87(23)$ & 11.08/7 & 0.1350 & 0.0176 & $\{c_{3},c_{4}\}$ & $19.08(31)$ & 7.714/6 & 0.2598 & 0.0575 \\
$\{c_{4},c_{5}\}$ & $18.789(61)$ & 11.19/7 & 0.1304 & 0.0166 & $\{c_{4},c_{5}\}$ & $18.905(76)$ & 6.59/6 & 0.3604 & 0.1008 \\
$\{c_{5},c_{6}\}$ & $18.784(48)$ & 11.22/7 & 0.1295 & 0.0164 & $\{c_{4},c_{6}\}$ & $18.94(11)$ & 6.96/6 & 0.3245 & 0.0837 \\
\hline
\end{tabular}
\caption{\label{tab:lambda_max_0.8_0.7}
Same as Tab.~\ref{tab:lambda_max_1.5_1.2}, but for $\hat{\lambda}_0^\mathrm{max}=0.8$ (left) and $\hat{\lambda}_0^\mathrm{max}=0.7$ (right).}
\end{table}

\begin{table}[!h]
\centering
\begin{tabular}{@{\hskip 5pt}l@{\hskip 5pt}c@{\hskip 5pt}c@{\hskip 5pt}c@{\hskip 5pt}c@{\hskip 5pt}|@{\hskip 5pt}l@{\hskip 5pt}c@{\hskip 5pt}c@{\hskip 5pt}c@{\hskip 5pt}c@{\hskip 5pt}}
\hline
fit ansatz & $f_\mathrm{c}$ & $\chi^2/\mathrm{dof}$ & $p$ & $w_\mathrm{AIC}$ & fit ansatz & $f_\mathrm{c}$ & $\chi^2/\mathrm{dof}$ & $p$ & $w_\mathrm{AIC}$ \\
\hline
linear & $18.805(23)$ & 6.058/7 & 0.5330 & 0.5892 & linear & $18.800(29)$ & 5.909/6 & 0.4335 & 0.3850 \\
\hline
$\{c_{2}\}$ & $18.836(91)$ & 5.906/6 & 0.4338 & 0.2338 & $\{c_{2}\}$ & $18.89(10)$ & 5.144/5 & 0.3985 & 0.2076 \\
$\{c_{3}\}$ & $18.86(33)$ & 6.033/6 & 0.4195 & 0.2194 & $\{c_{3}\}$ & $18.88(33)$ & 5.863/5 & 0.3198 & 0.1449 \\
$\{c_{4}\}$ & $18.807(26)$ & 5.936/6 & 0.4305 & 0.2304 & $\{c_{4}\}$ & $18.789(29)$ & 4.038/5 & 0.5439 & 0.3609 \\
$\{c_{5}\}$ & $18.823(55)$ & 5.861/6 & 0.4390 & 0.2392 & $\{c_{5}\}$ & $18.846(58)$ & 4.968/5 & 0.4197 & 0.2267 \\
$\{c_{6}\}$ & $18.823(58)$ & 5.901/6 & 0.4344 & 0.2345 & $\{c_{6}\}$ & $18.856(66)$ & 4.853/5 & 0.4341 & 0.2402 \\
\hline
\end{tabular}
\caption{\label{tab:lambda_max_0.6_0.5}
Same as Tab.~\ref{tab:lambda_max_1.5_1.2}, but for $\hat{\lambda}_0^\mathrm{max}=0.6$ (left) and $\hat{\lambda}_0^\mathrm{max}=0.5$ (right).}
\end{table}

\begin{table}[!h]
\centering
\begin{tabular}{@{\hskip 5pt}l@{\hskip 5pt}c@{\hskip 5pt}c@{\hskip 5pt}c@{\hskip 5pt}c@{\hskip 5pt}|@{\hskip 5pt}l@{\hskip 5pt}c@{\hskip 5pt}c@{\hskip 5pt}c@{\hskip 5pt}c@{\hskip 5pt}}
\hline
fit ansatz & $f_\mathrm{c}$ & $\chi^2/\mathrm{dof}$ & $p$ & $w_\mathrm{AIC}$ & fit ansatz & $f_\mathrm{c}$ & $\chi^2/\mathrm{dof}$ & $p$ & $w_\mathrm{AIC}$ \\
\hline
linear & $18.831(40)$ & 3.881/5 & 0.5667 & 0.6437 & linear & $18.813(49)$ & 3.68/4 & 0.4510 & 0.4316 \\
\hline
$\{c_{2}\}$ & $18.70(14)$ & 3.232/4 & 0.5198 & 0.3276 & $\{c_{2}\}$ & $18.52(29)$ & 2.773/3 & 0.4279 & 0.2499 \\
$\{c_{3}\}$ & $18.48(37)$ & 3.268/4 & 0.5139 & 0.3217 & $\{c_{3}\}$ & $18.33(66)$ & 3.187/3 & 0.3637 & 0.2032 \\
$\{c_{4}\}$ & $18.90(22)$ & 3.779/4 & 0.4367 & 0.2492 & $\{c_{4}\}$ & $18.75(49)$ & 3.665/3 & 0.3000 & 0.1600 \\
$\{c_{5}\}$ & $18.778(63)$ & 3.232/4 & 0.5197 & 0.3275 & $\{c_{5}\}$ & $18.715(89)$ & 2.316/3 & 0.5094 & 0.3141 \\
$\{c_{6}\}$ & $18.757(82)$ & 3.255/4 & 0.5161 & 0.3238 & $\{c_{6}\}$ & $18.63(16)$ & 2.432/3 & 0.4878 & 0.2965 \\
\hline
\end{tabular}
\caption{\label{tab:lambda_max_0.4_0.3}
Same as Tab.~\ref{tab:lambda_max_1.5_1.2}, but for $\hat{\lambda}_0^\mathrm{max}=0.4$ (left) and $\hat{\lambda}_0^\mathrm{max}=0.3$ (right).}
\end{table}

\begin{table}[!h]
\centering
\begin{tabular}{@{\hskip 5pt}l@{\hskip 5pt}c@{\hskip 5pt}c@{\hskip 5pt}c@{\hskip 5pt}c@{\hskip 5pt}|@{\hskip 5pt}l@{\hskip 5pt}c@{\hskip 5pt}c@{\hskip 5pt}c@{\hskip 5pt}c@{\hskip 5pt}}
\hline
fit ansatz & $f_\mathrm{c}$ & $\chi^2/\mathrm{dof}$ & $p$ & $w_\mathrm{AIC}$ & fit ansatz & $f_\mathrm{c}$ & $\chi^2/\mathrm{dof}$ & $p$ & $w_\mathrm{AIC}$ \\
\hline
linear & $18.744(58)$ & 0.6343/3 & 0.8885 & 1.2006 & linear & $18.81(10)$ & 0.1058/2 & 0.9485 & 0.9485 \\
\hline
$\{c_{2}\}$ & $19.00(42)$ & 0.2807/2 & 0.8691 & 0.5271 & $\{c_{2}\}$ & $18.76(76)$ & 0.1017/1 & 0.7498 & 0.3496 \\
$\{c_{3}\}$ & $19.24(85)$ & 0.3058/2 & 0.8582 & 0.5205 & $\{c_{3}\}$ & $18.8(15)$ & 0.1045/1 & 0.7465 & 0.3492 \\
$\{c_{4}\}$ & $19.02(51)$ & 0.3396/2 & 0.8438 & 0.5118 & $\{c_{4}\}$ & $18.78(77)$ & 0.1040/1 & 0.7471 & 0.3492 \\
$\{c_{5}\}$ & $18.80(11)$ & 0.2525/2 & 0.8814 & 0.5346 & $\{c_{5}\}$ & $18.81(11)$ & 0.09283/1 & 0.7606 & 0.3512 \\
$\{c_{6}\}$ & $18.89(24)$ & 0.2598/2 & 0.8782 & 0.5327 & $\{c_{6}\}$ & $18.78(40)$ & 0.09791/1 & 0.7544 & 0.3503 \\
\hline
\end{tabular}
\caption{\label{tab:lambda_max_0.25_0.2}
Same as Tab.~\ref{tab:lambda_max_1.5_1.2}, but for $\hat{\lambda}_0^\mathrm{max}=0.25$ (left) and $\hat{\lambda}_0^\mathrm{max}=0.2$ (right).}
\end{table}


\begin{table}[!h]
\centering
\begin{tabular}{@{\hskip 5pt}l@{\hskip 5pt}c@{\hskip 5pt}c@{\hskip 5pt}c@{\hskip 5pt}c@{\hskip 5pt}|@{\hskip 5pt}l@{\hskip 5pt}c@{\hskip 5pt}c@{\hskip 5pt}c@{\hskip 5pt}c@{\hskip 5pt}}
\hline
fit ansatz & $\Delta\langle\phi^2\rangle_\mathrm{c}/\lambda$ & $\chi^2/\mathrm{dof}$ & $p$ & $w_\mathrm{AIC}$ & fit ansatz & $\Delta\langle\phi^2\rangle_\mathrm{c}/\lambda$ & $\chi^2/\mathrm{dof}$ & $p$ & $w_\mathrm{AIC}$ \\
\hline
linear & $-0.003086(71)$ & 9.356/12 & 0.6722 & 1.3798 & linear & $-0.003125(95)$ & 8.968/11 & 0.6249 & 1.0162 \\
\hline
$\{c_{2}\}$ & $-0.00299(26)$ & 9.233/11 & 0.6004 & 0.5398 & $\{c_{2}\}$ & $-0.00274(32)$ & 7.494/10 & 0.6781 & 0.7811 \\
$\{c_{3}\}$ & $-0.00307(18)$ & 9.350/11 & 0.5896 & 0.5092 & $\{c_{3}\}$ & $-0.00371(53)$ & 7.789/10 & 0.6495 & 0.6741 \\
$\{c_{4}\}$ & $-0.00309(14)$ & 9.355/11 & 0.5892 & 0.5079 & $\{c_{4}\}$ & $-0.00298(16)$ & 7.747/10 & 0.6535 & 0.6884 \\
$\{c_{5}\}$ & $-0.00277(23)$ & 7.027/11 & 0.7969 & 1.6263 & $\{c_{5}\}$ & $-0.00273(30)$ & 6.989/10 & 0.7265 & 1.0055 \\
$\{c_{6}\}$ & $-0.00306(18)$ & 9.333/11 & 0.5912 & 0.5136 & $\{c_{6}\}$ & $-0.00289(21)$ & 7.536/10 & 0.6741 & 0.7649 \\
\hline
\end{tabular}
\caption{\label{tab:lambda_max_1.2_1.0_phisq}
Same as Tab.~\ref{tab:lambda_max_1.8_1.5_phisq}, but for $\hat{\lambda}_0^\mathrm{max}=1.2$ (left) and $\hat{\lambda}_0^\mathrm{max}=1.0$ (right).}
\end{table}


\begin{table}[!h]
\centering
\begin{tabular}{@{\hskip 5pt}l@{\hskip 5pt}c@{\hskip 5pt}c@{\hskip 5pt}c@{\hskip 5pt}c@{\hskip 5pt}|@{\hskip 5pt}l@{\hskip 5pt}c@{\hskip 5pt}c@{\hskip 5pt}c@{\hskip 5pt}c@{\hskip 5pt}}
\hline
fit ansatz & $\Delta\langle\phi^2\rangle_\mathrm{c}/\lambda$ & $\chi^2/\mathrm{dof}$ & $p$ & $w_\mathrm{AIC}$ & fit ansatz & $\Delta\langle\phi^2\rangle_\mathrm{c}/\lambda$ & $\chi^2/\mathrm{dof}$ & $p$ & $w_\mathrm{AIC}$ \\
\hline
linear & $-0.00312(11)$ & 8.967/10 & 0.5352 & 0.6166 & linear & $-0.00306(13)$ & 7.956/9 & 0.5386 & 0.6200 \\
\hline
$\{c_{2}\}$ & $-0.00259(38)$ & 6.682/9 & 0.6701 & 0.7109 & $\{c_{2}\}$ & $-0.00261(43)$ & 6.672/8 & 0.5724 & 0.4335 \\
$\{c_{3}\}$ & $-0.00485(90)$ & 5.802/9 & 0.7596 & 1.1042 & $\{c_{3}\}$ & $-0.0058(17)$ & 5.331/8 & 0.7217 & 0.8476 \\
$\{c_{4}\}$ & $-0.00290(17)$ & 6.057/9 & 0.7342 & 0.9717 & $\{c_{4}\}$ & $-0.00286(18)$ & 5.751/8 & 0.6751 & 0.6870 \\
$\{c_{5}\}$ & $-0.00272(30)$ & 6.851/9 & 0.6526 & 0.6534 & $\{c_{5}\}$ & $-0.00275(31)$ & 6.683/8 & 0.5712 & 0.4311 \\
$\{c_{6}\}$ & $-0.00276(24)$ & 6.263/9 & 0.7134 & 0.8769 & $\{c_{6}\}$ & $-0.00273(28)$ & 6.220/8 & 0.6226 & 0.5435 \\
\hline
\end{tabular}
\caption{\label{tab:lambda_max_0.9_0.8_phisq}
Same as Table~\ref{tab:lambda_max_1.8_1.5_phisq}, for $\hat{\lambda}_0^\mathrm{max}=0.9$ (left) and $\hat{\lambda}_0^\mathrm{max}=0.8$ (right).}
\end{table}


\clearpage


\section{Primary data\label{sec:app_d}}

The primary data for all our lattices are given in Tabs.~\ref{tab:finite_lattice_mu0sq}, \ref{tab:finite_lattice_phisq}; entries marked with a cross were not simulated.

\begin{table}[!hb]
    \centering
    \begin{tabular}{c|c@{\hskip 2pt}c@{\hskip 2pt}c@{\hskip 2pt}c@{\hskip 2pt}c@{\hskip 2pt}c}
    \hline
    $\hat{\lambda}_0$ & 52 & 64 & 80 & 112 & 136 & 156 \\
    \hline
    1.80 & $-1.136317(20)$ & $-1.136132(15)$ & $-1.135998(13)$ & $-1.135911(13)$ & $-1.1358874(78)$ & $-1.1358936(85)$ \\
    1.50 & $-0.964770(24)$ & $-0.964545(19)$ & $-0.9644496(74)$ & $-0.964354(14)$ & $-0.964320(26)$ & $-0.964295(13)$ \\
    1.20 & $-0.788141(25)$ & $-0.787916(16)$ & $-0.787798(12)$ & $-0.7876957(65)$ & $-0.7876715(55)$ & $-0.7876441(70)$ \\
    1.00 & $-0.667122(12)$ & $-0.6669207(93)$ & $-0.6667997(69)$ & $-0.6666679(72)$ & $-0.6666276(74)$ & $-0.6666221(79)$ \\
    0.90 & $-0.605534(15)$ & $-0.605311(10)$ & $-0.6051800(89)$ & $-0.6050353(62)$ & $-0.6050036(65)$ & $-0.6049914(66)$ \\
    0.80 & $-0.543089(16)$ & $-0.542855(10)$ & $-0.5427121(95)$ & $-0.5425974(58)$ & $-0.5425476(58)$ & $-0.5425410(55)$ \\
    0.70 & $-0.4797444(92)$ & $-0.4795280(77)$ & $-0.4793735(60)$ & $-0.4792619(48)$ & $-0.4792203(60)$ & $-0.4792016(58)$ \\
    0.60 & $-0.415456(43)$ & $-0.415215(32)$ & $-0.415103(23)$ & $-0.414958(19)$ & $-0.414909(10)$ & $-0.414902(10)$ \\
    0.50 & $-0.3500728(70)$ & $-0.3498497(54)$ & $-0.3497085(52)$ & $-0.3495691(43)$ & $-0.3495246(51)$ & $-0.3495035(49)$ \\
    0.40 & $-0.283532(15)$ & $-0.283272(16)$ & $-0.2831319(64)$ & $-0.2829844(71)$ & $-0.2829417(54)$ & $-0.2829170(60)$ \\
    0.30 & $-0.2155900(51)$ & $-0.2153594(44)$ & $-0.2152055(49)$ & $-0.2150682(37)$ & $-0.2150226(44)$ & $-0.2149901(50)$ \\
    0.25 & $-0.181052(18)$ & $-0.180818(16)$ & $-0.1806483(82)$ & $-0.1805269(48)$ & $-0.1804667(59)$ & $-0.1804548(65)$ \\
    0.20 & $-0.1460796(76)$ & $-0.1458560(65)$ & $-0.1456907(46)$ & $-0.1455513(46)$ & $-0.1454942(57)$ & $-0.1454815(61)$ \\
    0.15 & $-0.110637(16)$ & $-0.110390(14)$ & $-0.1102430(78)$ & $-0.1101073(53)$ & $-0.110069(10)$ & $-0.1100292(25)$ \\
    0.10 & $-0.0746293(74)$ & $-0.0744107(62)$ & $-0.0742685(43)$ & $-0.0741302(40)$ & $-0.0740775(53)$ & $-0.0740550(26)$ \\
    0.07 & $-0.0527239(80)$ & $-0.0525250(68)$ & $-0.0523814(31)$ & $-0.0522516(19)$ & $-0.0522057(24)$ & $-0.0521815(25)$ \\
    \hline
    $\hat{\lambda}_0$ & 176 & 216 & 256 & 320 & 344 & 368 \\
    \hline
    1.80 & $-1.135864(13)$ & x & x & x & x & x \\
    1.50 & $-0.964280(12)$ & x & x & x & x & x \\
    1.20 & $-0.7876435(92)$ & x & x & x & x & x \\
    1.00 & $-0.6666007(83)$ & x & x & x & x & x \\
    0.90 & $-0.6049714(66)$ & x & x & x & x & x \\
    0.80 & $-0.5425223(73)$ & x & x & x & x & x \\
    0.70 & $-0.4791849(48)$ & x & x & x & x & x \\
    0.60 & $-0.414889(11)$ & x & x & x & x & x \\
    0.50 & $-0.3494929(45)$ & $-0.3494762(35)$ & $-0.3494771(43)$ & x & x & x \\
    0.40 & $-0.2829104(71)$ & $-0.282902(14)$ & $-0.2828831(25)$ & x & x & x \\
    0.30 & $-0.2149789(51)$ & $-0.2149662(64)$ & $-0.2149483(47)$ & x & x & x \\
    0.25 & $-0.1804359(30)$ & $-0.1804180(33)$ & $-0.1804114(21)$ & x & x & x \\
    0.20 & $-0.1454638(66)$ & $-0.1454475(44)$ & $-0.1454340(50)$ & x & x & x \\
    0.15 & $-0.1100207(60)$ & $-0.1099990(27)$ & $-0.1099888(22)$ & $-0.1099803(11)$ & $-0.1099781(11)$ & x \\
    0.10 & $-0.0740357(33)$ & $-0.0740151(43)$ & $-0.0740086(17)$ & $-0.0739985(11)$ & $-0.0739969(11)$ & x \\
    0.07 & $-0.0521583(22)$ & $-0.0521424(17)$ & $-0.0521295(13)$ & $-0.05211991(93)$ & $-0.05211787(89)$ & $-0.0521163(10)$ \\
    \hline
    \end{tabular}
    \caption{\label{tab:finite_lattice_mu0sq}Survey of the pseudo-critical couplings $\hat{\mu}^2_{0\mathrm{c}}(\hat{\lambda}_0,N_x)$ in finite volume, obtained from the point where the Binder-cumulant reaches the universal value $U_4^*=0.46548$ given in Ref.~\cite{Ferrenberg:2018zst}, for all simulated lattices.}
\end{table}

\begin{table}[!hp]
    \centering
    \begin{tabular}{c|c@{\hskip 2pt}c@{\hskip 2pt}c@{\hskip 2pt}c@{\hskip 2pt}c@{\hskip 2pt}c}
    \hline
    $\hat{\lambda}_0$ & 52 & 64 & 80 & 112 & 136 & 156 \\
    \hline
    1.80 & $0.250937(13)$ & $0.250403(10)$ & $0.2499872(85)$ & $0.2495899(98)$ & $0.2494503(61)$ & $0.2493795(71)$ \\
    1.50 & $0.251625(19)$ & $0.251021(15)$ & $0.2505875(60)$ & $0.250154(12)$ & $0.249987(24)$ & $0.249885(12)$ \\
    1.20 & $0.252402(28)$ & $0.251743(19)$ & $0.251236(14)$ & $0.2507418(87)$ & $0.2505459(76)$ & $0.250439(10)$ \\
    1.00 & $0.253049(15)$ & $0.252317(12)$ & $0.2517505(98)$ & $0.251186(11)$ & $0.250962(12)$ & $0.250863(13)$ \\
    0.90 & $0.253424(20)$ & $0.252639(15)$ & $0.252039(14)$ & $0.251433(10)$ & $0.251201(11)$ & $0.251091(12)$ \\
    0.80 & $0.253847(24)$ & $0.252977(16)$ & $0.252324(16)$ & $0.251683(10)$ & $0.251444(11)$ & $0.251318(11)$ \\
    0.70 & $0.254276(12)$ & $0.2533709(99)$ & $0.2526539(75)$ & $0.2519723(67)$ & $0.2517183(83)$ & $0.2515692(84)$ \\
    0.60 & $0.254791(80)$ & $0.253834(66)$ & $0.253077(49)$ & $0.252307(44)$ & $0.252000(24)$ & $0.251839(25)$ \\
    0.50 & $0.255416(12)$ & $0.2543125(98)$ & $0.2534577(84)$ & $0.2526135(67)$ & $0.2523159(90)$ & $0.2521269(92)$ \\
    0.40 & $0.256243(31)$ & $0.254918(26)$ & $0.253997(14)$ & $0.252999(14)$ & $0.252651(12)$ & $0.252443(13)$ \\
    0.30 & $0.257320(13)$ & $0.2558000(96)$ & $0.254663(12)$ & $0.253541(11)$ & $0.253095(12)$ & $0.252856(14)$ \\
    0.25 & $0.258062(49)$ & $0.256344(44)$ & $0.255051(40)$ & $0.253847(16)$ & $0.253330(17)$ & $0.253095(20)$ \\
    0.20 & $0.259034(27)$ & $0.257146(26)$ & $0.255624(20)$ & $0.254221(18)$ & $0.253603(21)$ & $0.253384(25)$ \\
    0.15 & $0.260398(72)$ & $0.258152(57)$ & $0.256389(35)$ & $0.254661(22)$ & $0.254128(30)$ & $0.2537100(90)$ \\
    0.10 & $0.262768(54)$ & $0.259804(43)$ & $0.257637(30)$ & $0.255489(34)$ & $0.254657(35)$ & $0.254231(20)$ \\
    0.07 & $0.26526(13)$ & $0.26169(15)$ & $0.258965(27)$ & $0.256299(14)$ & $0.255297(20)$ & $0.254774(20)$ \\
    \hline
    $\hat{\lambda}_0$ & 176 & 216 & 256 & 320 & 344 & 368 \\
    \hline
    1.80 & $0.249311(10)$ & x & x & x & x & x \\
    1.50 & $0.249807(10)$ & x & x & x & x & x \\
    1.20 & $0.250371(14)$ & x & x & x & x & x \\
    1.00 & $0.250771(14)$ & x & x & x & x & x \\
    0.90 & $0.251000(12)$ & x & x & x & x & x \\
    0.80 & $0.251219(15)$ & x & x & x & x & x \\
    0.70 & $0.2514631(76)$ & x & x & x & x & x \\
    0.60 & $0.251767(29)$ & x & x & x & x & x \\
    0.50 & $0.2520232(89)$ & $0.2518531(73)$ & $0.2517431(79)$ & x & x & x \\
    0.40 & $0.252318(17)$ & $0.252144(28)$ & $0.252021(18)$ & x & x & x \\
    0.30 & $0.252681(15)$ & $0.252477(17)$ & $0.252355(14)$ & x & x & x \\
    0.25 & $0.2529016(90)$ & $0.252674(10)$ & $0.2525358(72)$ & x & x & x \\
    0.20 & $0.253155(23)$ & $0.2528859(51)$ & $0.2527207(49)$ & x & x & x \\
    0.15 & $0.253493(13)$ & $0.2531537(56)$ & $0.2529471(59)$ & $0.2527633(55)$ & $0.2527175(58)$ & x \\
    0.10 & $0.253907(21)$ & $0.2535197(61)$ & $0.253264(11)$ & $0.2530154(73)$ & $0.2529650(73)$ & x \\
    0.07 & $0.254355(26)$ & $0.2538521(86)$ & $0.2535299(98)$ & $0.253229(11)$ & $0.2531679(88)$ & $0.253099(10)$ \\
    \hline
    \end{tabular}
    \caption{\label{tab:finite_lattice_phisq}$\langle\phi^2\rangle_\mathrm{c}(\hat{\lambda}_0,N_x)$, i.e., $\langle\phi^2\rangle$ evaluated at $\hat{\mu}_{0\mathrm{c}}^2(\hat{\lambda}_0,N_x)$ as determined in Tab.~\ref{tab:finite_lattice_mu0sq}, for all simulated lattices.}
\end{table}


\clearpage



\end{document}